\documentclass[journal]{IEEEtran}

\usepackage{times}
\usepackage{epsfig}
\usepackage{graphicx}

\usepackage{xcolor,soul,framed} 

\graphicspath{{../pdf/}{../jpeg/}}
\DeclareGraphicsExtensions{.pdf,.jpeg,.png}
\usepackage{epstopdf}
\usepackage[cmex10]{amsmath}
\usepackage{array}
\usepackage{mdwmath}
\usepackage{mdwtab}
\usepackage{eqparbox}
\usepackage{url}
\usepackage{amssymb}
\usepackage{amsmath}
\usepackage{booktabs}
\usepackage{multirow}
\usepackage{algorithm}
\usepackage{algorithmic}
\usepackage{makecell}
\graphicspath{{figure/}}
\newcommand{\tensor}[1]{\boldsymbol{\mathcal{#1}}}

\newcommand{\mat}[1]{\mathbf{#1}}
\newcommand{\vect}[1]{\mathbf{#1}}

\usepackage{enumitem}
\newtheorem{theorem}{Theorem}

\newtheorem{myDef}{Definition}

\newtheorem{proof}{Proof}[section]
\newtheorem{proposition}[theorem]{Proposition}
\newtheorem{remark}{Remark}[section]

\ifCLASSINFOpdf

\else

\fi

\usepackage[pagebackref=true,breaklinks=true,letterpaper=true,colorlinks,bookmarks=false]{hyperref}

\begin{document}
\title{Hyperspectral Super-Resolution via Coupled Tensor Ring Factorization}


\author{
Wei~He$^{1}$, Yong Chen$^{2}$, Naoto~Yokoya$^1$,  Chao~Li$^1$, Qibin~Zhao$^1$\\
$^1$RIKEN AIP\ \ \ $^2$UESTC \\
\tt\small \{wei.he;naoto.yokoya;chao.li;qibin.Zhao\}@riken.jp, chenyong1872008@163.com
}
\maketitle

\begin{abstract}
   Hyperspectral super-resolution (HSR) fuses a low-resolution hyperspectral image (HSI) and a high-resolution multispectral image (MSI) to obtain a high-resolution HSI (HR-HSI). 
   In this paper, we propose a new model, named coupled tensor ring factorization (CTRF), for HSR.
   The proposed CTRF approach simultaneously learns high spectral resolution core tensor from the HSI and high spatial resolution core tensors from the MSI, and reconstructs the HR-HSI via tensor ring (TR) representation (Figure~\ref{fig:framework}).
   The CTRF model can separately exploit the low-rank property of each class (Section \ref{sec:analysis}), which has been never explored in the previous coupled tensor model. Meanwhile,
   it inherits the simple representation of coupled matrix/CP factorization and flexible low-rank exploration of coupled Tucker factorization.
   Guided by Theorem~\ref{th:1}, we further propose a spectral nuclear norm regularization to explore the global spectral low-rank property.
   The experiments have demonstrated the advantage of the proposed nuclear norm regularized CTRF (NCTRF) as compared to previous matrix/tensor and deep learning methods.
\end{abstract}
\vspace{-7pt}
\IEEEpeerreviewmaketitle
\section{Introduction}


Benefit from the wealthy spectral information, hyperspectral image (HSI) has been widely used in different remote sensing and computer version applications~\cite{ghassemian2016review,li2017pixel,shen2016integrated}. However, due to the limitation of imaging techniques~\cite{kanatsoulis2018hyperspectral,yokoya2017hyperspectral}, there is a trade-off between spatial and spectral resolutions~\cite{dian2017hyperspectral,Shaw2003}. The hyperspectral imaging sensors are hard to obtain high resolution HSIs (HR-HSI). Meanwhile, the multispectral imaging sensors always sacrifice the spectral resolution to obtain the high-resolution multispectral images (MSI)~\cite{wei2016multiband,yokoya2017hyperspectral}. Therefore, the topic of \textit{hyperspectral super-resolution} (HSR) or \textit{hyperspectral and multispectral image fusion}, which fuses low-resolution HSI and high-resolution MSI to generate a HR-HSI~\cite{kawakami2011high,yokoya2017hyperspectral} is important and attractive in real-would applications.\par


Hyperspectral super-resolution (HSR) has a long history~\cite{vivone2015critical}. Initially, HSR tries to fuse the HSI and a panchromatic image~\cite{loncan2015hyperspectral}, including multiresolution analysis (MRA)~\cite{vivone2015critical} and sparse representation based methods~\cite{chen2015sirf}. However, these methods are limited to enhance the spatial details in practice~\cite{yokoya2017hyperspectral}. Subsequently, A Bayesian framework has been introduced to obtain a HR-HSI from the HSI and MSI, \textit{i.e.,} maximum a posteriori (MAP)~\cite{hardie2004map,wei2015fast} and Bayesian sparse representation~\cite{akhtar2015bayesian}. Very recently, deep learning has also been introduced to fuse the HSI and MSI, and has achieved remarkable results~\cite{dian2018deep,qu2018unsupervised,xie2019multispectral,shi2018deep,nie2018deeply}. However, deep learning related methods always need numerous training samples, which is problematic for the real HSR in the remote sensing society. In this paper, we will focus on the unsupervised matrix/tensor related methods. \par
\begin{figure}[!t]
	\centering
	\includegraphics[width= 1.0 \linewidth]{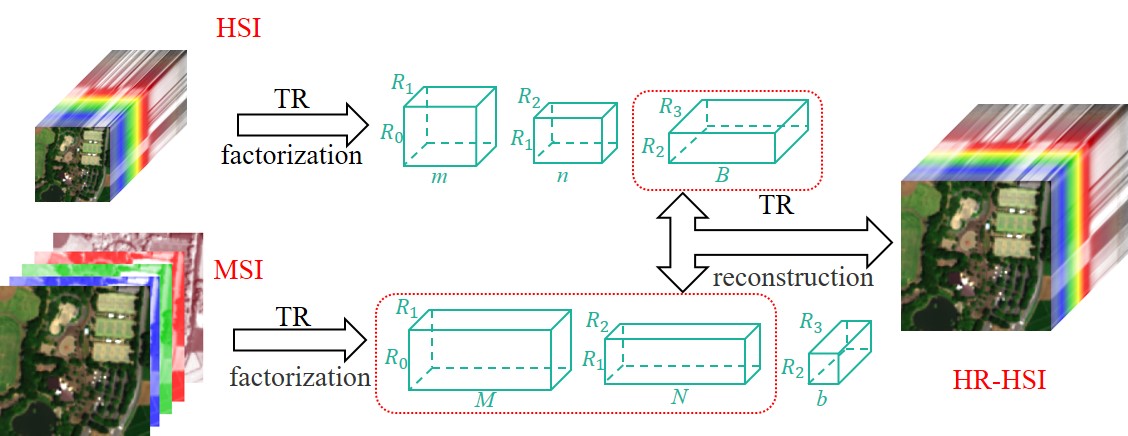}
	\caption{
    Illustration of coupled tensor ring factorization.
	}
	\label{fig:framework}
	\vspace{-15pt}
\end{figure}
Over the past few years, numerous state-of-the-art HSR methods are achieved by matrix factorization models~\cite{akhtar2015bayesian,dian2019learning,dong2016hyperspectral,kawakami2011high,lanaras2015hyperspectral,simoes2015convex,veganzones2016hyperspectral,wei2015hyperspectral,wei2016multiband,wei2015fast,wycoff2013non,XuTGRS2019,yokoya2012coupled}. The idea is to reshape the original HR-HSI into the matrix, and decompose the matrix to the basis and  coefficient matrices. The coupled matrix factorization model tries to jointly learn the spectral basis from the factorization of HSI, and the spatial coefficients from the MSI. Generally, the unfolding operator to reshape the 3-D HR-HSI to the 2-D matrix results in the loss of spatial correlation. Although several strategies, \textit{e.g.,} non-negative regularization of basis and coefficients~\cite{dong2016hyperspectral, yokoya2012coupled} and total variation regularization of coefficient matrix~\cite{simoes2015convex} have been introduced to improve the HSR quality, the spatial-spectral correlations have not been fully exploited. \par
\begin{figure}[!t]
	\centering
	\includegraphics[width= 1.0 \linewidth]{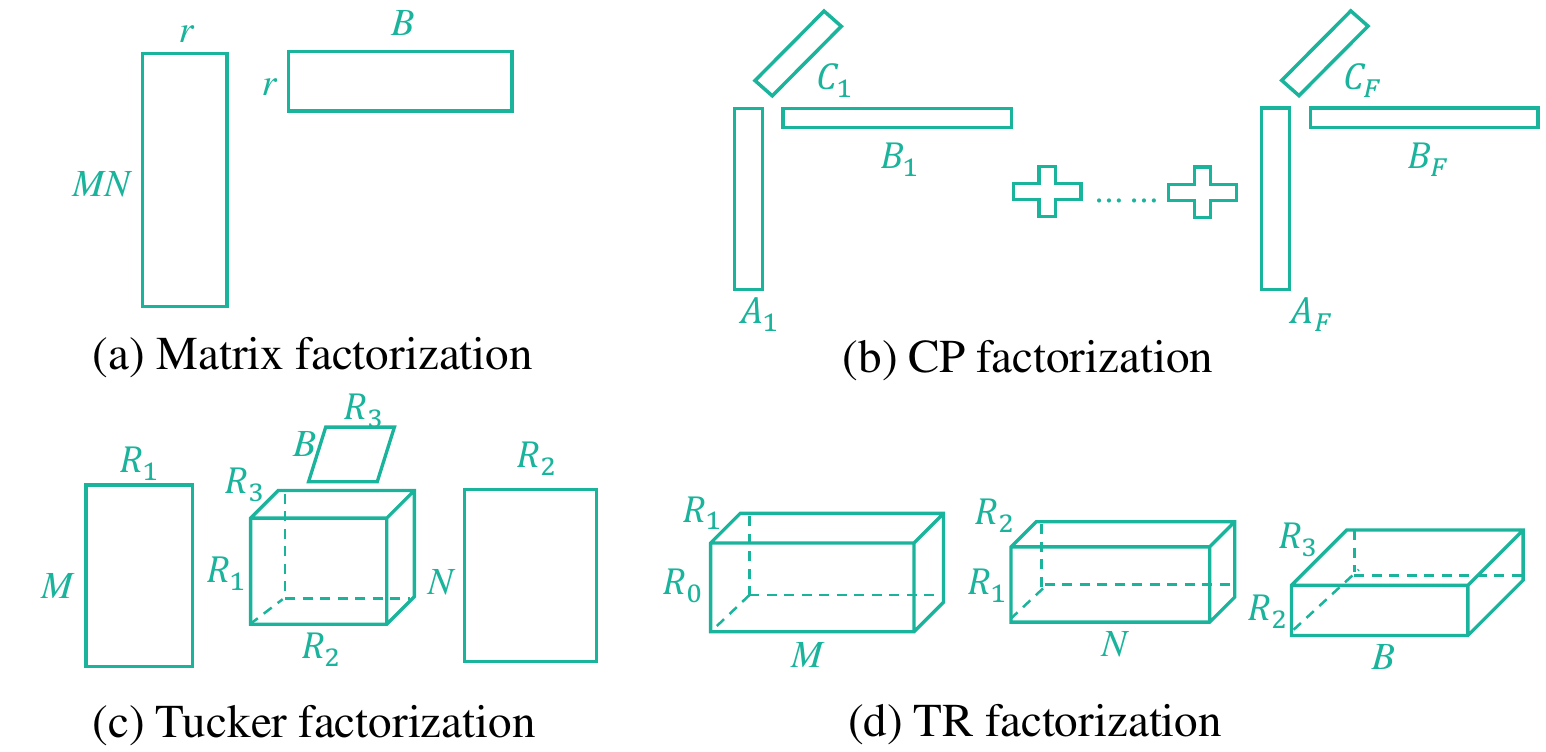}
	\caption{
    Matrix/tensor factorization of a 3-D tensor of size $\mathbb{R}^{M\times N \times B}$. For (a), the tensor is reshaped to the matrix.
	}
	\vspace{-7pt}
	\label{fig:Decom}
\end{figure}

To extend the coupled matrix factorization model, a coupled tensor factorization model has recently been developed to the HSR. The well-known approaches are a coupled canonical polyadic (CP) factorization model~\cite{kanatsoulis2018hyperspectral} and a coupled Tucker factorization model~\cite{li2018fusing,li2018coupled}. Figure~\ref{fig:Decom} illustrates the different matrix/tensor factorization models of a 3-D tensor. As presented in \cite{kanatsoulis2018hyperspectral}, the CP factorization model assumes that the low-rank properties of different dimensions are the same; however, it is not true in HR-HSI~\cite{chang2017hyper,yokoya2017hyperspectral}. The Tucker decomposition model introduces a 3-D core tensor, which is independent of the whole spatial and spectral degradation processing. The existence of this core tensor increases the complexity of the model for computation and estimation. Several strategies, \textit{i.e.,} $L_1$ norm~\cite{li2018fusing} and $L_2$ norm ~\cite{li2018coupled} regularizations of the core tensor, and non-local processing~\cite{dian2017hyperspectral,xu2019nonlocal} have been proposed to improve the accuracy. However, the improvement is limited.\par
In this paper, we propose a coupled tensor ring factorization (CTRF) model for HSR.
Tensor ring (TR) factorization~\cite{zhao2016tensor} tries to decompose a high-order tensor to a series of 3-D tensors, called TR cores.
From Figure~\ref{fig:Decom}, it can be observed that the structure of tensor ring (TR) \cite{wang2017efficient,zhao2016tensor} factorization inherits the simple representation of matrix/CP factorization, and flexible low-rank exploration of Tucker factorization. Furthermore, we illustrate in Section~\ref{sec:analysis} that the TR factorization model can exploit the low-rank property of different classes, which has been never explored in the previous coupled tensor factorization models.
Thereby, we separately apply TR factorization to HSI and MSI, and obtain a high spectral resolution core tensor from HSI, meanwhile, two high spatial resolution core tensors from MSI. Subsequently, the HR-HSI is reconstructed from the high resolution core tensors via TR. The illustration is presented in Figure~\ref{fig:framework}.
The TR factorization model ignores the global low-rank property of original HR-HSI along the spectral dimension, which is proved important in HSR~\cite{yokoya2017hyperspectral,HE2016TGRS}. Fortunately, we find that the spectral low-rank property of HR-HSI is bounded by the rank of third TR core along mode-2 unfolding (Theorem \ref{th:1}). It motivates us to develop a nuclear norm regularized CTRF (NCTRF) model. The main contributions of this paper are summarized as follows:

\begin{itemize}[leftmargin = 10pt]
\item We propose a coupled tensor ring factorization model for the HSR task. The nuclear norm regularization of the third TR core with mode-2 unfolding is proposed to further exploit the global spectral low-rank property of HR-HSI.

\item We analysis the superiority of CTRF for HSR, and develop an efficient alternating iteration method for the proposed model. The experiments demonstrate the advantage of NCTRF model compared to previous matrix/tensor and deep learning methods.

\end{itemize}
\noindent
\textbf{Notations.}
\label{notation}
We mainly adopt the notations from \cite{kolda2009tensor} in this paper. Tensors of order $N\geq 3$ are denoted by boldface Euler script letters, e.g., $\tensor{H}\in\mathbb{R}^{I_1\times I_2\times\cdots \times I_N}$. Scalars are denoted by normal lowercase letters or uppercase letters, e.g., $h, H \in\mathbb{R}$. $H(i_1,\cdots,i_N)$ denotes the element of tensor $\tensor{H}$ in position $(i_1,\cdots,i_N)$. Vectors are denoted by boldface lowercase letters, e.g., $\vect{h}\in\mathbb{R}^{I}$. Matrices are denoted by boldface capital letters, e.g., $\mat{H}\in\mathbb{R}^{I\times J}$. Moreover, we employ two types of tensor unfolding (matricization) operations in this paper. The first mode-$n$ unfolding \cite{kolda2009tensor} of tensor $\tensor{H}  \in\mathbb{R}^{I_1\times I_2\times\cdots \times I_N}$ is denoted by $\mat{H}_{(n)}\in\mathbb{R}^{I_n \times  {I_1 \cdots I_{n-1} I_{n+1} \cdots I_N}}$. The second mode-$n$ unfolding of tensor $\tensor{H}$ which is often used in TR operations \cite{zhao2016tensor} is denoted by $\mat{H}_{<n>}\in\mathbb{R}^{I_n \times  {I_{n+1} \cdots I_{N} I_{1} \cdots I_{n-1}}}$. $\mat{H}[I_1 \cdots I_i,I_{i+1} \cdots I_N]$ is the matricization by regarding the first $i$ dimensions as row and the last $N-i$ dimensions as column. 
We define the folding operation for the first mode-$n$ unfolding as $\text{fold}_n(\cdot)$, i.e., for a tensor $\tensor{H}$, we have $\text{fold}_n(\mat{H}_{(n)})=\tensor{H}$. In addition, the inner product of two tensors $\tensor{H}$, $\tensor{W}$ with the same size $\mathbb{R}^{I_1\times I_2\times\cdots \times I_N}$ is defined as $\langle \tensor{H},\tensor{W} \rangle=\sum_{i_1}\sum_{i_2}\cdots\sum_{i_N}h_{i_1 i_2\ldots i_N}w_{i_1 i_2\ldots i_N}$. Furthermore, the Frobenius norm of $\tensor{H}$ is defined by $\left \| \tensor{H} \right \|_F=\sqrt{\langle \tensor{H},\tensor{H} \rangle}$. \par
In this paper, we adopt $ \tensor{X} \in \mathbb{R}^{M \times N \times B}$, $ \tensor{Y} \in \mathbb{R}^{m \times n \times B}$, $ \tensor{Z} \in \mathbb{R}^{M \times N \times b}$ to stand for the registered HR-HSI, HSI and MSI, respectively. Here, $M > n$ and $N>n$ represents the spatial size and $B>b$ stands for the spectral size. The objective of super-resolution is to estimate HR-HSI by the fusion of HSI and MSI.

\section{Related Works}
In this section, we briefly introduce the related coupled matrix/tensor decomposition HSR methods.

\vspace{-4pt}
\subsection{Coupled matrix factorization}
The coupled matrix factorization based methods have been contributed by various studies~\cite{yokoya2012coupled,simoes2015convex,wei2015fast,wei2016multiband,wei2015hyperspectral,veganzones2016hyperspectral,wycoff2013non,dian2019learning,akhtar2015bayesian,lanaras2015hyperspectral,kawakami2011high}. Specifically, HR-HSI is characterized by strong spectral correlation among different pixels and can be expressed as the following matrix factorization model,
\vspace{-3px}
\begin{equation}
\label{CMF1}
\mat{X}_{(3)} = \mat{C}\mat{D},
\end{equation}
where $\mat{X}_{(3)}  \in  \mathbb{R}^{B \times MN}$ stands for the unfolding HR-HSI tensor $\tensor{X}$ along the spectral dimension, $\mat{C}\in \mathbb{R}^{B \times L}$ and  $\mat{D}\in \mathbb{R}^{L \times MN}$ represent the spectral basis and the related corresponding coefficient matrix, respectively. In the factorization model \eqref{CMF1}, the basis matrix can be allocated with different physical meanings with different regularizations, \textit{i.e.,} low dimensional subspace with low-rank regularization~\cite{simoes2015convex,wei2015fast,kawakami2011high}, and spectral endmember matrix with non-negative regularization~\cite{wei2016multiband,yokoya2012coupled,wycoff2013non}. In order to explore the relationship between HR-HSI, HSI and MSI, the researchers usually assume that there exist two linear degradation matrices $\mat{P}_0 \in  \mathbb{R}^{mn \times MN}$, $\mat{P}_3 \in  \mathbb{R}^{b \times B}$ such that
\vspace{-3px}
\begin{equation}
\label{CMF2}
\mat{Y}_{(3)} = \mat{C}(\mat{D}\mat{P}_0^{\top}), \ \ \mat{Z}_{(3)} = (\mat{P}_3 \mat{C})\mat{D}.
\end{equation}
Here, $\mat{P}_0$ is adopted to modelling the point spread function (PSF)~\cite{yokoya2017hyperspectral}. The basic idea of coupled matrix factorization model for the HR-HSI approaches is to jointly factorize $\mat{D}$ and $\mat{C}$ from HSI and MSI, and reconstruct the HR-HSI via equation \eqref{CMF1}.

However, the coupled matrix factorization based methods need to reshape the 2-D dimension spatial image into 1-D vector, ignoring the spatial correlation of the image. Total variation has been introduced to enhance the spatial smoothness of the corresponding coefficient matrix $\mat{D}$~\cite{simoes2015convex}. Unfortunately, the related total variation regularized methods introduce the spatial oversmooth, and the accuracy of super-resolution result is limited~\cite{kanatsoulis2018hyperspectral}. \par

\subsection{Coupled CP factorization}
The CP decomposition~\cite{kolda2009tensor,sidiropoulos2017tensor} tries to decompose the tensor $\tensor{X}$ into the sum of multiple rank-1 tensors, which can be expressed as
\vspace{-3px}
\begin{equation}
\label{CP1}
\tensor{X} = \sum_{f=1}^{F}  \vect{a}_f \circ \vect{b}_f \circ \vect{c}_f,
\end{equation}
where $\circ$ stands for the outer product, and $F$ is the \textit{rank} of CP decomposition. $\mat{A} = [\vect{a}_1,\ldots,\vect{a}_f]$, $\mat{B} = [\vect{b}_1,\ldots,\vect{b}_f]$, $\mat{C} = [\vect{c}_1,\ldots,\vect{c}_f]$ are called the  low-rank latent factors and CP decomposition can be represented as $\tensor{X} = \left \lfloor \mat{A},\mat{B},\mat{C} \right \rfloor$. \par
The coupled CP factorization model~\cite{kanatsoulis2018hyperspectral} assumes that there exist three linear degradation matrices $\mat{P}_1 \in  \mathbb{R}^{m \times M}$, $\mat{P}_2 \in  \mathbb{R}^{n \times N}$, $\mat{P}_3 \in  \mathbb{R}^{b \times B}$ such that
\vspace{-3px}
\begin{equation}
\label{CP2}
\tensor{Y} = \left \lfloor \mat{P}_1\mat{A},\mat{P}_2\mat{B},\mat{C} \right \rfloor, \tensor{Z} = \left \lfloor \mat{A},\mat{B},\mat{P}_3\mat{C} \right \rfloor
\end{equation}
The advantage of the coupled CP factorization model is to preserve the HR-HSI spatial structure. However, HR-HSI has stronger spectral low-rank property compared to spatial perspective. In the coupled CP model, each latent factors $\mat{A}$, $\mat{B}$, $\mat{C}$ are treated equally with a much larger \textit{rank} $F$, resulting in the insufficient spectral low-rank exploration.

\subsection{Coupled Tucker factorization}
The Tucker decomposition~\cite{tucker1966some,kolda2009tensor} decomposes the tensor $\tensor{X}$ into
\vspace{-3px}
\begin{equation}
\label{Tuck1}
\tensor{X} = \tensor{O} \times_1 \mat{A} \times_2 \mat{B} \times_3 \mat{C},
\end{equation}
where
$\tensor{O} \in  \mathbb{R}^{R_1 \times R_2 \times R_3}$ stands for the core tensor, and $\mat{A} \in \mathbb{R}^{M \times R_1}, \mat{B}  \in  \mathbb{R}^{N \times R_2}, \mat{C}  \in  \mathbb{R}^{B \times R_3}$ are the factors with related to different dimensions. \par

The coupled Tucker factorization~\cite{li2018coupled,li2018fusing} also assumes the degradation matrices $\mat{P}_1 \in  \mathbb{R}^{m \times M}$, $\mat{P}_2 \in  \mathbb{R}^{n \times N}$, $\mat{P}_3 \in  \mathbb{R}^{b \times B}$ such that
\vspace{-3px}
\begin{equation}
\label{Tuck2}
\begin{aligned}
\tensor{Y} = \tensor{O} \times_1 (\mat{P}_1 \mat{A}) \times_2 (\mat{P}_2 \mat{B}) \times_3 \mat{C},\\
\tensor{Z} = \tensor{O} \times_1 \mat{A} \times_2 \mat{B} \times_3 (\mat{P}_3 \mat{C}).
\end{aligned}
\end{equation}

In the coupled Tucker factorization model, the factor size can be chosen according to the low-rank prior of each dimension. However, it introduces a core tensor $\tensor{O}$ which is independent of the whole spatial and spectral degradation matrices. $L_1$ norm~\cite{li2018fusing} and $L_2$ norm~\cite{li2018coupled} regularizers have been introduced to help the identification of $\tensor{O}$; however, the related regularizations introduce additional turning parameters, and the accurate estimation of core tensor from HSI and MSI remains a problem~\cite{li2018fusing}.

\section{Coupled Tensor Ring Factorization}
In this section, we propose a CTRF model for HSR. The proposed CTRF model has a more flexible \textit{rank} selection strategy compared to CP decomposition, and a simpler representation compared to Tucker decomposition.

\subsection{Tensor ring decomposition}
\label{TR_sec}
TR decomposition represents the tensor $\tensor{H}$ by circular multilinear products over a sequence of three-order core tensors $\tensor{G}:=\{\tensor{G}^{(1)},\ldots,\tensor{G}^{(N)}\}$, where $\tensor{G}^{(n)} \in\mathbb{R}^{R_{n} \times I_{n} \times R_{n+1}}$, $n=1,2,\cdots,N$, $R_1=R_{N+1}$~\cite{zhao2016tensor}. Here, $R = [R_1,\cdots,R_N]$ is the TR rank. Each element of the tensor $\tensor{H}$ can be rewritten as
\vspace{-3px}
\begin{equation}
\begin{split}
\label{TRD}
H(i_1,\cdots,i_N) =
\mathrm{Tr}(\mat{G}^{(1)}(i_1)\cdots\mat{G}^{(N)}(i_N)),
\end{split}
\end{equation}
where $\mat{G}^{(n)}(i_n)$ is the $i_n$-th lateral slice matrix of the core tensor $\tensor{G}^{(n)}$. In this paper, we adopt the notation $\tensor{H} = \Phi(\tensor{G})$ to represent the TR decomposition. The illustration of TR decomposition can be found in Figure~\ref{fig:Decom}. Next, we introduce two important properties of TR decomposition. \par

\begin{figure*}[!t]
  \centering
  \begin{minipage}[t]{0.24\textwidth}\centering
    \includegraphics[width=\textwidth]{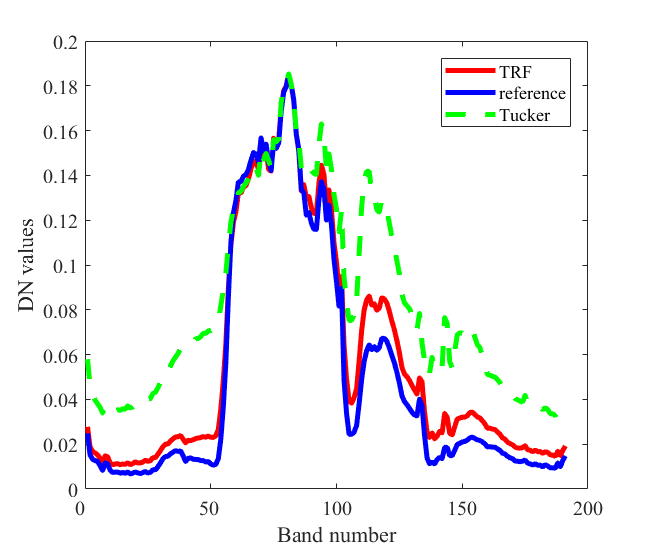} \\ \footnotesize (a) class1-sigature1
    \end{minipage}
    \begin{minipage}[t]{0.24\textwidth}\centering
    \includegraphics[width=\textwidth]{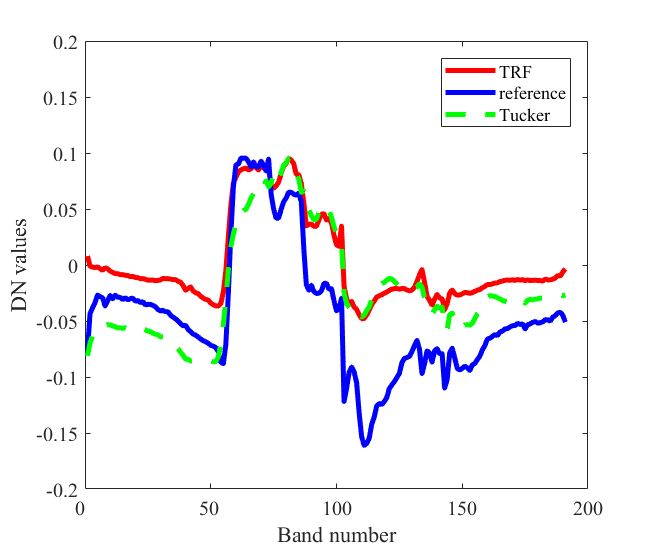} \\\footnotesize (b) class1-sigature2
    \end{minipage}
    \begin{minipage}[t]{0.24\textwidth}\centering
    \includegraphics[width=\textwidth]{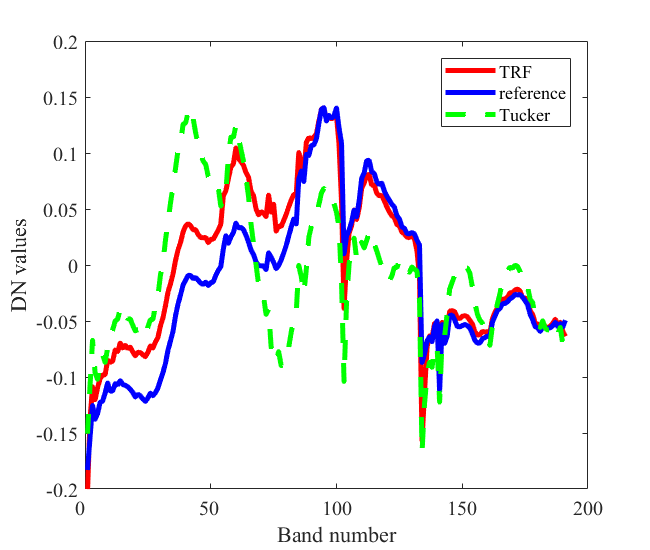} \\\footnotesize (c) class2-sigature1
    \end{minipage}
    \begin{minipage}[t]{0.24\textwidth}\centering
    \includegraphics[width=\textwidth]{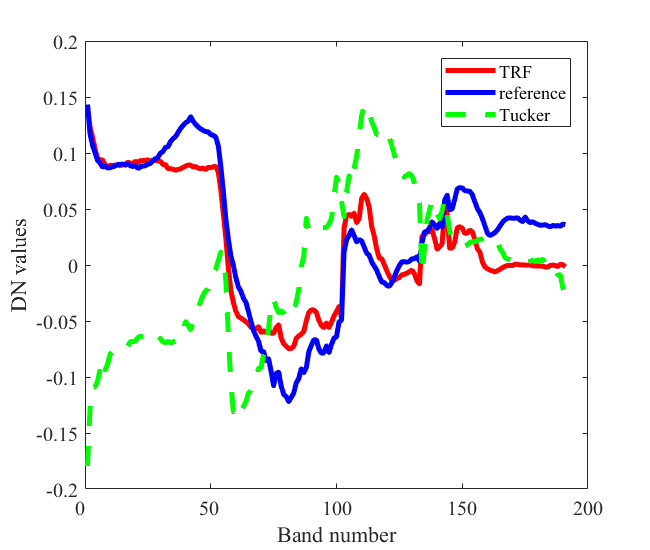} \\\footnotesize (d) class2-sigature2
    \end{minipage}
      \vspace{2px}
  \caption{The signature comparison between TR factorization (TRF), Tucker decomposition and the reference. The reference means the signature extracted from each class via SVD. Class1-sigature1 means the first signature from class 1.}
  \vspace{-7px}
\label{fig:motivation}
\end{figure*}

\begin{myDef}(Multilinear product~\cite{zhao2018learning}.) \label{pr:1}
Suppose $\tensor{G}^{(n)}$ and $\tensor{G}^{(n+1)}$ are the two nearby cores of TR decomposition. The multilinear product of the two cores is $\tensor{G}^{(n,n+1)}\in\mathbb{R}^{R_{n}\times I_nI_{n+1}\times R_{n+2}}$ and denoted by
\vspace{-3px}
\begin{equation}
    \begin{split}
    \label{multipro}
     \tensor{G}^{(n,n+1)}((j_l-1)I_n+i_k) = \tensor{G}^{(n)}(i_k)\tensor{G}^{(n+1)}(j_l),
    \end{split}
    \end{equation}
    for $i_k=1,\cdots,I_n, \ j_l=1,\cdots,I_{n+1}$.
\end{myDef}

From Definition~\ref{pr:1}, if the tensor $\tensor{H}$ can be decomposed via \eqref{TRD}, we can immediately obtain the following
\vspace{-3px}
\begin{equation}
\label{mat:trans}
\mat{H}[I_1\cdots I_n,I_{n+1}\cdots I_N] =
\mat{G}^{(1,\cdots,n)}_{(2)}\times (\mat{G}^{(n+1,\cdots,N)}_{<2>})^{\top},
\end{equation}
where $\mat{G}_{(2)}^{(1,\cdots,n)}$ denotes the multilinear products of the first $n$ cores and the unfolding along mode-2, and $\mat{G}^{(n+1,\cdots,N)}_{<2>}$ represents the multilinear products of the last $N-n$ cores and the second unfolding along mode-2. That is to say, the rank value $R_1\times{}R_{n+1}$ is bounded by the rank of unfolding matrix $\mat{H}[I_1\cdots I_n,I_{n+1}\cdots I_N]$.

\begin{proposition}(Circular dimensional permutation invariance ~\cite{zhao2016tensor}.) \label{pr:2}
 If $\overleftarrow{\tensor{H}_n} \in \mathbb{R}^{I_{n+1}\times\cdots\times I_{N}\times I_1 \times{}\cdots \times I_n} $ is denoted as the circularly
shifting the dimensions of $\tensor{H}$ by $n$, then we have $\overleftarrow{\tensor{H}_n}=\Phi(\{\tensor{G}^{(n+1)},\ldots,\tensor{G}^{(N)},\tensor{G}^{(1)},\ldots,\tensor{G}^{(n)}\})$
\end{proposition}

With Proposition~\ref{pr:2}, we can easily shift the middle cores of TR to the first position, and utilize Definition~\ref{pr:1} to analysis each core separately.

\subsection{Coupled tensor ring factorization}
Analogously to coupled CP and Tucker factorization, we also assume the spatial degradation matrices $\mat{P}_1 \in  \mathbb{R}^{m \times M}$, $\mat{P}_2 \in  \mathbb{R}^{n \times N}$, and spectral degradation matrix $\mat{P}_3 \in  \mathbb{R}^{b \times B}$. $\mat{P}_1$ and $\mat{P}_2$ can be regarded as the separable operators of $\mat{P}_0$ in the coupled matrix factorization model, as $\mat{P}_0 =\mat{P}_1 \otimes \mat{P}_2 $, with $\otimes$ standing for the Kronecker product. As discussed in ~\cite{rivenson2009compressed}, the separable operators have more advantage in optimization and calculation. With this in mind, the TR decomposition of HR-HSI can be represented as
\vspace{-3px}
\begin{equation}
\label{TR1}
\tensor{X} = \mat{\Phi}(\tensor{G}^{(1)},\tensor{G}^{(2)},\tensor{G}^{(3)}),
\end{equation}
where $\tensor{G}^{(1)} \in\mathbb{R}^{R_{1} \times {M} \times R_{2}} $, $\tensor{G}_{(2)} \in\mathbb{R}^{R_{2} \times {N} \times R_{3}} $ and $\tensor{G}^{(3)} \in\mathbb{R}^{R_{3} \times {B} \times R_{1}} $ stand for the core tensors related to spatial, spatial and spectral dimensions, respectively. $R = [R_1,R_2,R_3]$ represents the TR \textit{rank} of HR-HSI. The HSI can be expressed as
\vspace{-3px}
\begin{equation}
\label{TR2}
\tensor{Y} = \mat{\Phi}(\tensor{G}^{(1)} \times_2 \mat{P}_1,\tensor{G}^{(2)} \times_2 \mat{P}_2,\tensor{G}^{(3)}),
\end{equation}
and the MSI can be formulated as
\vspace{-3px}
\begin{equation}
\label{TR3}
\tensor{Z} = \mat{\Phi}(\tensor{G}^{(1)},\tensor{G}^{(2)},\tensor{G}^{(3)} \times_2 \mat{P}_3).
\end{equation}
By combining \eqref{TR2} and \eqref{TR3}, we can obtain our CTRF model as
\vspace{-3px}
\begin{align}
&
\min_{\tensor{G}^{(1)},\tensor{G}^{(2)},\tensor{G}^{(3)}} \left \| \tensor{Y} - \mat{\Phi}(\tensor{G}^{(1)} \times_2 \mat{P}_1,\tensor{G}^{(2)} \times_2 \mat{P}_2,\tensor{G}^{(3)}) \right \|_F^2
\notag
\\
&
+\left \| \tensor{Z} - \mat{\Phi}(\tensor{G}^{(1)},\tensor{G}^{(2)},\tensor{G}^{(3)} \times_2 \mat{P}_3) \right \|_F^2.
\label{TR4}
\end{align}

\subsection{Motivation of CTRF for HSR}
\label{sec:analysis}
In this section, we look into \eqref{TR4}, and see the insights why the proposed CTRF model can beat the coupled Tucker decomposition model in HSR. \par
For HSIs, each spectral pixel, denoted as $ \tensor{X}(i,j,:)$ of the size $\mathbb{R}^B$,
stands for the spectrum of one specific class. Typically, one scene of HSI contains multiple classes, and there are several basis spectral signatures to construct the low-dimensional subspace for each class~\cite{BioucasTGRS2008}. Tucker decomposition usually reshapes HSIs along the spectral dimension, and exploits the spectral low rank property globally. From another side, TR factorization can learn the spectral core $\tensor{G}^{(3)} \in\mathbb{R}^{R_{3} \times {B} \times R_{1}} $ with $R_{3}$ standing for the number of classes, and $R_{1}$ representing the subspace dimension of each class. \par

An experiment is constructed to demonstrate the advantage of CTRF. We select $50$ pixels of class1 (grass) and $50$ pixels of class2 (road) from WDC dataset (Section \ref{database}), and construct these $100$ pixels as the HR-HSI of size $10 \times 10 \times 191$.
TR factorization can obtain a spectral core of size ${2 \times {191} \times 2} $ from the constructed HR-HSI. Figure~\ref{fig:motivation} illustrates the spectral signatures extracted by TR and Tucker decomposition. We apply singular value decomposition (SVD) to each class and extract the first two principal component vectors as the reference signatures. In Figure~\ref{fig:motivation}, the signatures obtained by TR are much more similar to those of references, indicating the advantage of TR factorization for HSI processing. \par

\subsection{NCTRF}
As analyzed in Section \ref{sec:analysis}, the obtained spectral core $\tensor{G}^{(3)}$ regards the subspace dimension of each class as the same. However, in the real case, the subspace dimensions of different classes may different. If we set a larger subspace dimension $R_{1}$, we will lose the global low-rank property of the spectral core tensor.
However, as reviewed in ~\cite{He_2019_CVPR,veganzones2016hyperspectral,wei2016multiband,yokoya2017hyperspectral}, the HSI has strong spectral correlation, indicating the low-rank property of HSI along the spectral dimension. Inspired by this fact, we propose to regularize the third core tensor $\tensor{G}^{(3)}$ along mode-2 as low rank, denoted as $rank(\tensor{G}^{(3)}_{<2>})$. The rank-constraint is hard to optimize, and we introduce the nuclear norm $\left \| \centerdot \right \|_*$, the sum of the singular values of the matrix~\cite{cai2010singular}, to regularize the low-rank property. Therefore, the proposed CTRF model with nuclear norm regularization (NCTRF) is formulated as
\begin{equation}
\begin{aligned}
\label{TR5}
\min_{\tensor{G}^{(1)},\tensor{G}^{(2)},\tensor{G}^{(3)}} \left \| \tensor{Y} - \mat{\Phi}(\tensor{G}^{(1)} \times_2 \mat{P}_1,\tensor{G}^{(2)} \times_2 \mat{P}_2,\tensor{G}^{(3)}) \right \|_F^2  \\ +
\left \| \tensor{Z} - \mat{\Phi}(\tensor{G}^{(1)},\tensor{G}^{(2)},\tensor{G}^{(3)} \times_2 \mat{P}_3) \right \|_F^2+\lambda \left \| \tensor{G}^{(3)}_{<2>} \right \|_*.
\end{aligned}
\end{equation}

\begin{theorem} \label{th:1}
Suppose an $N$-th order tensor $\tensor{H}\in\mathbb{R}^{I_1\times I_2\times\cdots \times I_N}$, we have the following property
\vspace{-3px}
\begin{equation}
\label{TR6}
rank(\tensor{G}^{(n)}_{<2>})\geq rank(\tensor{H}_{(n)}),
\end{equation}
for $n=1,\cdots N$.
\end{theorem}
By {Theorem~\ref{th:1}}, we can immediately conclude that the spectral rank of HR-HSI $\tensor{X}$ can be bounded by the rank of core tensor $\tensor{G}^{(3)}$ along mode-2. The regularization $\left \| \tensor{G}^{(3)}_{<2>} \right \|_*$ can exploit the global low-rank property along spectral dimension.

\begin{remark}
In this paper, we only regularize the spectral low-rank property, and ignore the spatial low-rank property, due to the reason that the low-rank property along the spectral dimension is much stronger than that of spatial dimension~\cite{chang2017hyper,HE2016TGRS}.
The NCTRF model~\eqref{TR5} introduces an additional parameter $\lambda$. To reduce the complexity of the proposed model, we fix the $\lambda$ as a constant value in the whole experiments. See supplementary material for more details.
\end{remark}

\subsection{Optimization}
The objective function \eqref{TR5} is non-convex. Fortunately, for each separable variable, the objective function is convex. We first introduce a latent variable $\tensor{G}^{(0)}= \tensor{G}^{(3)}$, and convert \eqref{TR5} to the following augmented Lagrangian function
\vspace{-3px}
\begin{align}
&
g(\tensor{G}^{(1)},\tensor{G}^{(2)},\tensor{G}^{(3)},\tensor{G}^{(0)},\tensor{L},\mu)=
\notag
\\
&
\left \| \tensor{Y} - \mat{\Phi}(\tensor{G}^{(1)} \times_2 \mat{P}_1,\tensor{G}^{(2)} \times_2 \mat{P}_2,\tensor{G}^{(3)}) \right \|_F^2
\notag
\\
&
+\left \| \tensor{Z} - \mat{\Phi}(\tensor{G}^{(1)},\tensor{G}^{(2)},\tensor{G}^{(3)} \times_2 \mat{P}_3) \right \|_F^2 +\lambda \left \| \tensor{G}^{(0)}_{(2)} \right \|_*
\notag
\\
&
+<\tensor{L},\tensor{G}^{(0)}- \tensor{G}^{(3)}> + \frac{\mu}{2}\left \| \tensor{G}^{(0)}- \tensor{G}^{(3)} \right \|_F^2,
\label{TRALM}
\end{align}
where $\tensor{L}$ represents the Lagrangian multiplier and $\mu$ stands for the penalty parameter. Next, we adopt an alternating iterative method~\cite{lin2011linearized} to optimize ~\eqref{TRALM}. \par
By fixing other variables, and update one variable for each iteration, the optimization of \eqref{TRALM} can be split into four subproblems.

Update $\tensor{G}^{(1)}$:
By fixing other variables, the update of $\tensor{G}^{(1)}$ can be formulated as
\vspace{-3px}
\begin{equation}
\begin{aligned}
\label{eq:up1}
\tensor{G}^{(1)}=\min_{\tensor{G}^{(1)}} g(\tensor{G}^{(1)},\tensor{G}^{(2)},\tensor{G}^{(3)},\tensor{G}^{(0)},\tensor{L},\mu).
\end{aligned}
\end{equation}
By using two kinds of tensor unfolding as in Section \ref{notation} and Definition~\ref{pr:1}, we can convert the optimization of \eqref{eq:up1} to the following problem
\vspace{-3px}
\begin{equation}
\label{eq:up11}
\begin{aligned}
\tensor{G}^{(1)}=\arg\min_{\tensor{G}^{(1)}} \left \| \mat{Y}_{<1>} -\mat{P}_1\mat{G}_{(2)}^{(1)}\mat{A}_1 \right \|_F^2 \\
+ \left \| \mat{Z}_{<1>} -\mat{G}_{(2)}^{(1)}\mat{B}_1 \right \|_F^2,
\end{aligned}
\end{equation}
where $\mat{Y}_{<1>}$ and $\mat{Z}_{<1>}$ stand for the second unfolding of the tensors $\tensor{Y}, \tensor{Z}$, respectively, $\mat{A}_1 = ((\tensor{G}^{(2)} \times _2 \mat{P}_2)\centerdot \tensor{G}^{(3)})_{<2>}^{\top}$ and $\mat{B}_1 = (\tensor{G}^{(2)} \centerdot (\tensor{G}^{(3)}\times _2 \mat{P}_3))_{<2>}^{\top}$. Optimization \eqref{eq:up11} is quadratic and its unique solution is equal
to solve the general Sylvester matrix equation~\cite{wei2015fast}
\vspace{-3px}
\begin{equation}
\label{eq:up111}
\begin{aligned}
\mat{P}_1^{\top} \mat{P}_1 \mat{G}_{(2)}^{(1)} \mat{A}_1\mat{A}_1^{\top} + \mat{G}_{(2)}^{(1)} \mat{B}_1\mat{B}_1^{\top}\\
 = \mat{P}_1^{\top} \mat{Y}_{<1>} \mat{A}_1^{\top} + \mat{Z}_{<1>}\mat{B}_1^{\top}.
\end{aligned}
\end{equation}
To avoid the large scale matrix inversion in the closed-form solution of \eqref{eq:up111}, we adopt conjugate gradient (CG) \cite{li2018fusing} to solve \eqref{eq:up111}.

Update $\tensor{G}^{(2)}$ and $\tensor{G}^{(3)}$: We firstly fix other variables and update $\tensor{G}^{(2)}$. We adopt Proposition~\ref{pr:2} to circularly shift the tensors $\tensor{Y}$ and $\tensor{Z}$, and convert the optimization of $\tensor{G}^{(2)}$ to the following optimization:
\vspace{-3px}
\begin{equation}
\begin{aligned}
\label{eq:up2}
\min_{\tensor{G}^{(2)}} \left \| \overleftarrow{\tensor{Y}_1} - \mat{\Phi}(\tensor{G}^{(2)} \times_2 \mat{P}_2,\tensor{G}^{(3)}, \tensor{G}^{(1)} \times_2 \mat{P}_1) \right \|_F^2  \\ +
\left \| \overleftarrow{\tensor{Z}_1} - \mat{\Phi}(\tensor{G}^{(2)},\tensor{G}^{(3)} \times_2 \mat{P}_3, \tensor{G}^{(1)}) \right \|_F^2.
\end{aligned}
\end{equation}
\eqref{eq:up2} can be convert to the following matrix version optimization problem
\vspace{-3px}
\begin{equation}
\label{eq:up22}
\begin{aligned}
\tensor{G}^{(2)}=\arg\min_{\tensor{G}^{(2)}} \left \| \mat{Y}_{<2>} -\mat{P}_2\mat{G}_{(2)}^{(2)}\mat{A}_2 \right \|_F^2 \\
+ \left \| \mat{Z}_{<2>} -\mat{G}_{(2)}^{(2)}\mat{B}_2 \right \|_F^2,
\end{aligned}
\end{equation}
where $\mat{Y}_{<2>}$ and $\mat{Z}_{<2>}$ are equal to the mode-1 unfolding of the tensors $\overleftarrow{\tensor{Y}_1}, \overleftarrow{\tensor{Z}_1}$, respectively, $\mat{A}_2 = (\tensor{G}^{(3)} \centerdot (\tensor{G}^{(1)}\times _2 \mat{P}_1))_{<2>}^{\top}$ and $\mat{B}_2 = ((\tensor{G}^{(3)}\times _2 \mat{P}_3) \centerdot \tensor{G}^{(1)})_{<2>}^{\top}$. Thus, it can also be efficiently solved by the CG method. \par
As to the update of $\tensor{G}^{(3)}$, we also adopt Proposition~\ref{pr:2} to circularly shift the tensors $\tensor{Y}$ and $\tensor{Z}$, and obtain the following optimization
\vspace{-3px}
\begin{equation}
\begin{aligned}
\label{eq:up3}
\min_{\tensor{G}^{(3)}} \left \| \overleftarrow{\tensor{Y}_2} - \mat{\Phi}(\tensor{G}^{(3)}, \tensor{G}^{(1)} \times_2 \mat{P}_1,\tensor{G}^{(2)} \times_2 \mat{P}_2) \right \|_F^2  \\ +
\left \| \overleftarrow{\tensor{Z}_2} - \mat{\Phi}(\tensor{G}^{(3)} \times_2 \mat{P}_3, \tensor{G}^{(1)},\tensor{G}^{(2)}) \right \|_F^2 \\
+<\tensor{L},\tensor{G}^{(0)}- \tensor{G}^{(3)}> + \frac{\mu}{2}\left \| \tensor{G}^{(0)}- \tensor{G}^{(3)} \right \|_F^2,
\end{aligned}
\end{equation}
which can be solved by the CG method.

Update $\tensor{G}^{(0)}$: By fixing other variables, the optimization of $\tensor{G}^{(0)}$ can be obtained by solving the following problem
\vspace{-3px}
\begin{equation}
\label{eq:up4}
\min_{\tensor{G}^{(0)}} \lambda \left \| \mat{G}^{(0)}_{(2)} \right \|_*
 + \frac{\mu}{2}\left \| \mat{G}_{(2)}^{(0)}- \mat{G}_{(2)}^{(3)} + \mat{L}_{(2)}/{\mu} \right \|_F^2,
\end{equation}
which can be solved by the closed-form solution~\cite{cai2010singular}
\vspace{-3px}
\begin{equation}
\label{eq:up44}
\tensor{G}^{(0)} = \text{fold}_2(\tensor{T}_{\frac{\lambda}{\mu}} (\mat{G}_{(2)}^{(3)}- \mat{L}_{(2)}/{\mu}) ).
\end{equation}
Here, $\tensor{T}_{\frac{\lambda}{\mu}}$ is the singular value thresholding (SVT) operator. \par

Update $\tensor{L}$ and $\mu$: Finally, we adopt the strategy in \cite{lin2011linearized} to update $\tensor{L}$ and $\mu$
\begin{equation}
\label{eq:up5}
\tensor{L} =\tensor{L} + \mu(\tensor{G}^{(0)}- \tensor{G}^{(3)}),\ \ \mu = min(\mu_{1},\rho \mu),
\end{equation}
where $\mu_1$ and $\rho>1$ denotes constant values. \par
To summary, the optimization to the proposed NCTRF model is presented in Algorithm~\ref{alg:NCTRF}.

\begin{algorithm}[!t]
\caption{Optimization of NCTRF}
\label{alg:NCTRF}
\begin{algorithmic}[1]
	\REQUIRE HSI $\tensor{Y}$, MSI $\tensor{Z}$, \textit{rank} $R$.
	\STATE $\lambda = 0.001, \rho = 1.5, \mu = 10^{-4}, \mu_1 = 10^{6}$
	\FOR{$i = 1, 2, 3, \cdots iter$}
	\STATE Update TR cores $\tensor{G}$ via \eqref{eq:up1}, \eqref{eq:up2}, \eqref{eq:up3}, \eqref{eq:up4}.
	\STATE Update  $\tensor{L}$ and $\mu$ via \eqref{eq:up5}.
	\\
	\ENDFOR
	
	\RETURN  HR-HSI $\tensor{X} = \mat{\phi}(\tensor{G})$;
\end{algorithmic}
\end{algorithm}

\begin{table}[!t]
  \centering
  \footnotesize
  \caption{Quantitative comparison of different algorithms under various noise levels. The results are the average of the three dataset, and best results are in bold. $\uparrow$ stands for the larger, the better, and $\downarrow$ is the inverse.}
    \begin{tabular}{cccccc}
    \hline
    \multicolumn{6}{c}{ SNR = 20} \\
    \hline
    Method & PSNR $\uparrow$  & RMSE$\downarrow$  & ERGAS$\downarrow$ & SAM$\downarrow$   & SSIM$\uparrow$  \\ \hline
    CNMF  & 26.73 & 12.86 & 4.38  & 8.53   & 0.752  \\
    FUSE  & 30.43 & 8.23  & 2.94  & 5.79   & 0.870  \\
    HySure & 27.56 & 11.72 & 4.51  & 9.64  & 0.727  \\
    STEREO & 30.31 & 8.20  & 3.06  & 7.35  & 0.846  \\
    CSTF  & 31.48 & 7.16  & 2.76  & 6.17   & 0.865  \\
    NLSTF & 25.47 & 14.36 & 5.42  & 12.99  & 0.688  \\
    CTRF  & 31.94 & 6.89  & 2.56  & 5.76   & 0.877  \\
    NCTRF & \textbf{31.98} & \textbf{6.84} & \textbf{2.55} & \textbf{5.74}  & \textbf{0.877} \\
    \hline
    \multicolumn{6}{c}{ SNR = 30} \\
    \hline
    CNMF  & 29.15 & 9.97  & 3.56  & 4.85   & 0.885 \\
    FUSE  & 35.63 & 4.42  & 1.67  & 3.30   & 0.956 \\
    HySure & 34.98 & 4.74  & 1.84  & 3.69  & 0.945  \\
    STEREO & 35.53 & 4.68  & 1.85  & 4.46  & 0.942 \\
    CSTF  & 37.93 & 3.47  & 1.32  & 3.21   & 0.963  \\
    NLSTF & 34.84 & 5.03  & 1.82  & 4.62   & 0.940 \\
    CTRF  & 38.15 & \textbf{3.46} & \textbf{1.31} & \textbf{3.27} & 0.965  \\
    NCTRF & \textbf{38.16} & \textbf{3.46} & \textbf{1.31} & \textbf{3.27}  & \textbf{0.966}  \\
    \hline
    \multicolumn{6}{c}{ SNR = 40} \\
    \hline
    CNMF  & 29.31 & 9.66  & 3.60  & 4.95   & 0.905  \\
    FUSE  & 37.61 & 3.50  & 1.40  & 2.48   & 0.980  \\
    HySure & 37.41 & 3.59  & 1.43  & 2.56  & 0.979 \\
    STEREO & 38.72 & 3.44  & 1.25  & 3.32  & 0.968  \\
    CSTF  & 40.69 & 2.58  & 1.02  & 2.45  & 0.981   \\
    NLSTF & \textbf{41.99} & \textbf{2.34} & \textbf{0.85} & \textbf{2.22}  & \textbf{0.985} \\
    CTRF  & 41.35 & 2.57  & 1.00  & 2.47   & \textbf{0.985}  \\
    NCTRF & 41.54 & 2.54  & 0.98  & 2.46   & \textbf{0.985}  \\
    \hline
    \end{tabular}%
  \label{tab:QEva}%
\vspace{-9pt}
\end{table}%

\section{Experiments}
In this section, we present the experimental results of different methods, followed by the parameter analysis, convergence analysis, computational time, and the comparison with deep learning related methods. The experiments are programmed in Matlab R2018b on a laptop with CPU Core i7-8750H 32G memory.

\subsection{Experimental database}
\label{database}
To validate the performance of our proposed NCTRF method for HSR, we conduct the experiments on three HR-HSI datasets, including Washington DC Mall (WDC)
\footnote{\url{https://engineering.purdue.edu/~biehl/MultiSpec/hyperspectral}},
Pavia Center (PaC)
\footnote{\url{http://www.ehu.eus/ccwintco/index.php/}}
and Indian Pines
\footnote{\url{https://engineering.purdue.edu/~biehl/MultiSpec/}}.
These three dataset are widely used in the simulation-based evaluation of HSR~\cite{kanatsoulis2018hyperspectral,li2018fusing}. The size of each HR-HSI, and the simulated HSIs and MSIs are presented in Table~\ref{tab:datasize}. We chose PSF with the average kernel (with size $8$) to generate the spatial degradation matrices. When we simulate HSI and MSI from HR-HSI, the Gaussion noise is added to the HSI and MSI, with signal-to-noise (SNR) ratio changes from 20dB, 30dB to 40dB. All the HR-HSIs are scaled to [0, 255].

\begin{table}[!t]
\footnotesize
\centering
\vspace{-5px}
\caption{The size of the image used for HSR experiments.}
\begin{tabular}{c | c | c |c}
	\hline
	  Image name  & HR-HSI                   & HSI                    & MSI        \\ \hline
	  WDC         & 256$\times$256$\times$90 & 64$\times$64$\times$90 & 256$\times$256$\times$4   \\ \hline
	  PaC         & 200$\times$200$\times$93 & 50$\times$50$\times$93 & 200$\times$200$\times$4   \\ \hline
	  Indian      & 144$\times$144$\times$100 & 16$\times$16$\times$100 & 144$\times$144$\times$4   \\ \hline
      CAVE        & 512$\times$512$\times$31 & 16$\times$16$\times$31 & 512$\times$512$\times$3   \\ \hline
\end{tabular}
\label{tab:datasize}
\vspace{-7px}
\end{table}

\begin{figure*}[!ht]
  \centering
    \begin{minipage}[t]{0.09\textwidth}\centering
        \includegraphics[width=\textwidth]{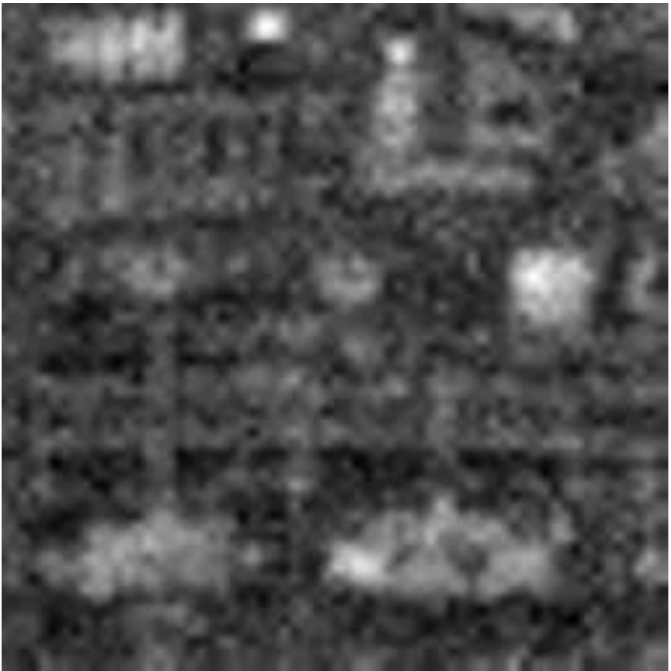} \\
    \includegraphics[width=\textwidth]{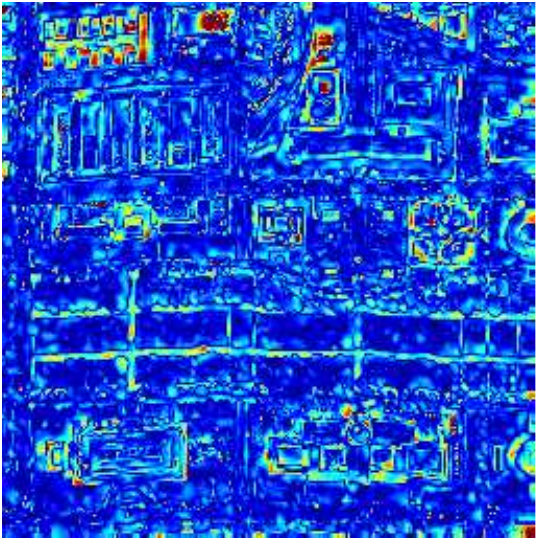} \\
    \end{minipage}
    \begin{minipage}[t]{0.09\textwidth}\centering
      \includegraphics[width=\textwidth]{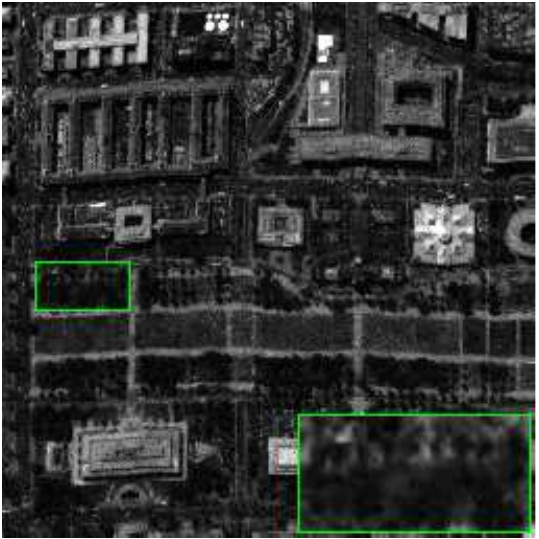} \\
    \includegraphics[width=\textwidth]{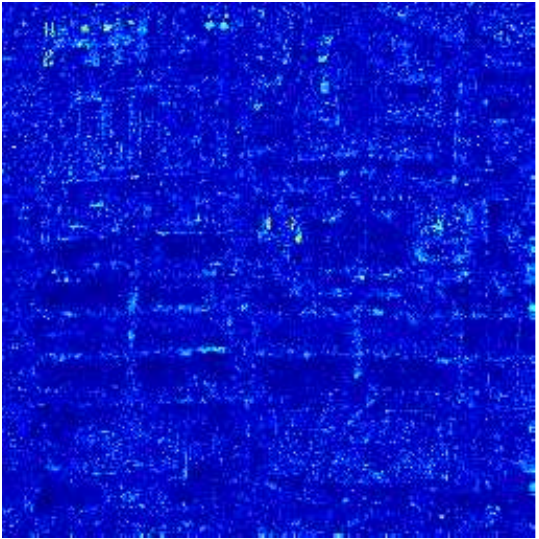} \\
    \end{minipage}
     \begin{minipage}[t]{0.09\textwidth}\centering
      \includegraphics[width=\textwidth]{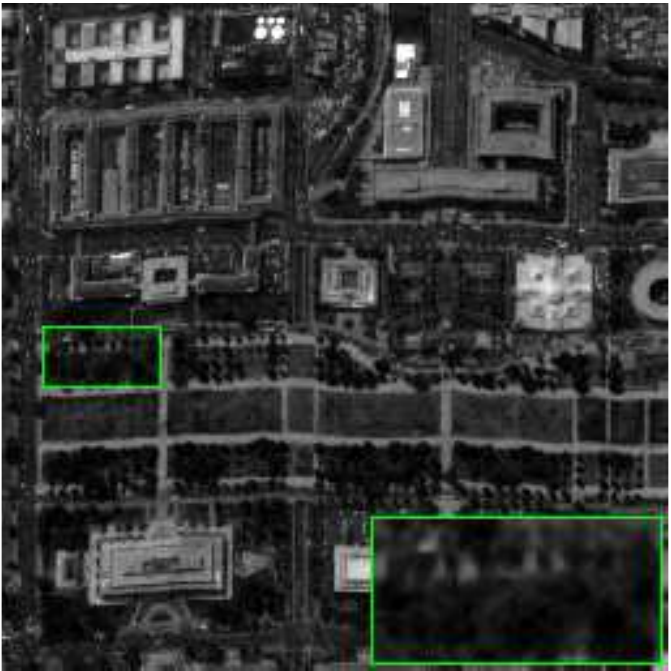} \\
    \includegraphics[width=\textwidth]{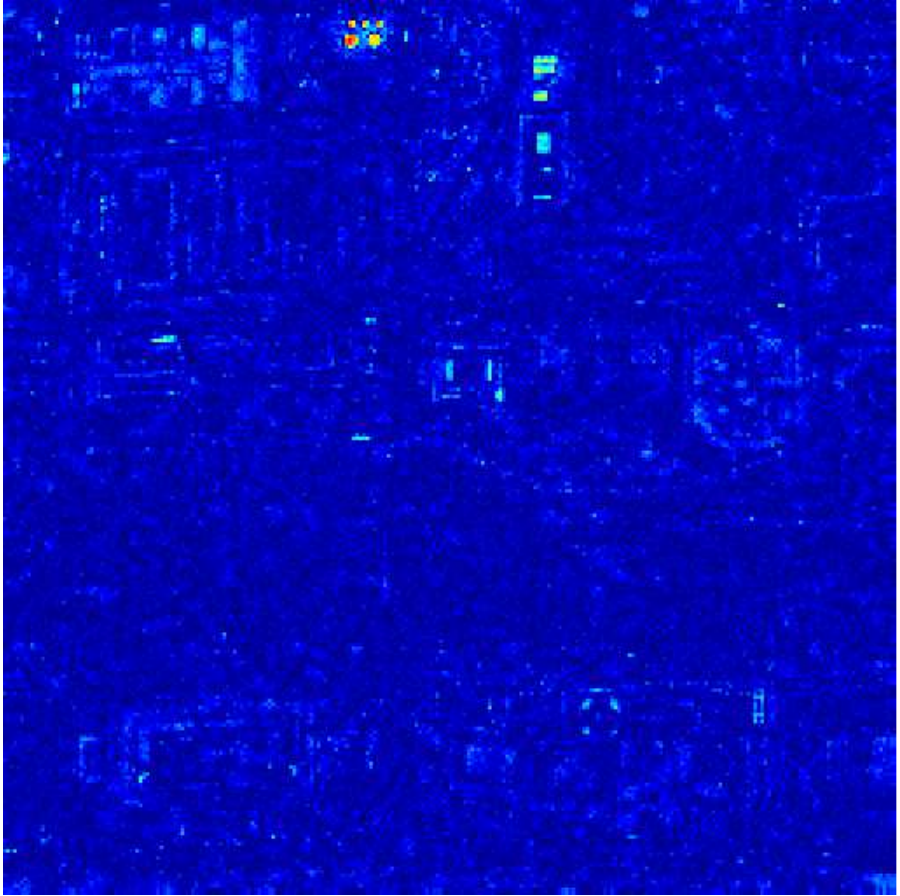} \\
    \end{minipage}
  \begin{minipage}[t]{0.09\textwidth}\centering
   \includegraphics[width=\textwidth]{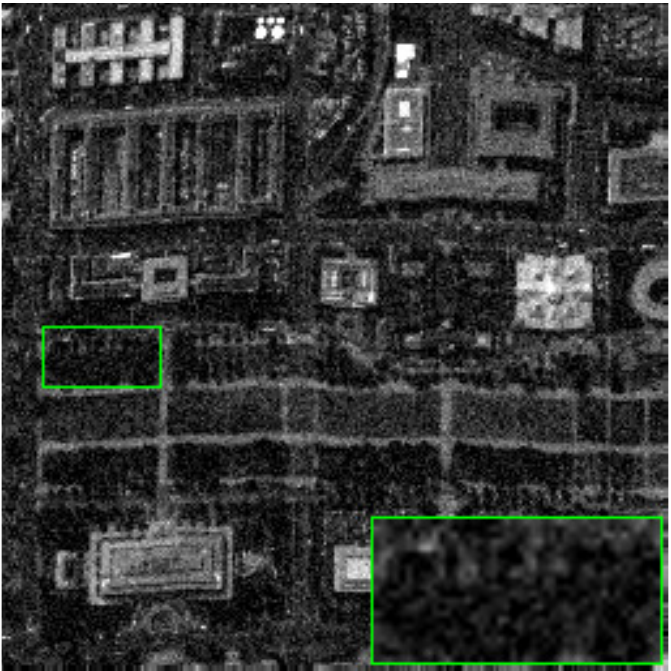} \\
    \includegraphics[width=\textwidth]{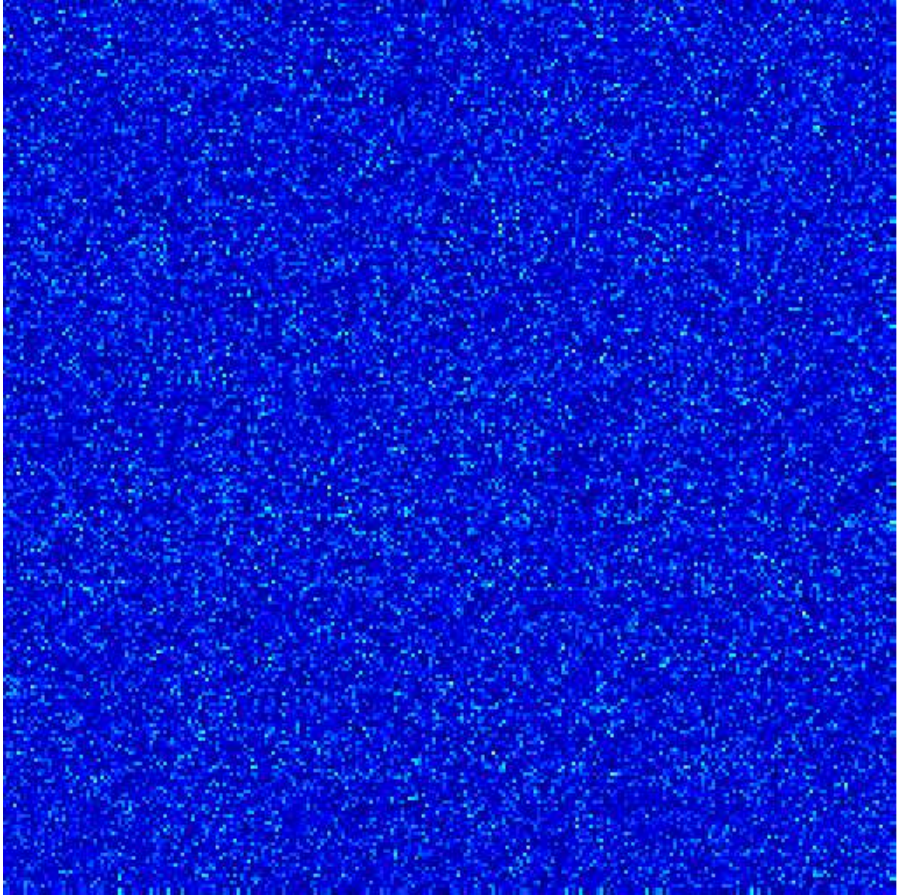} \\
    \end{minipage}
    \begin{minipage}[t]{0.09\textwidth}\centering
     \includegraphics[width=\textwidth]{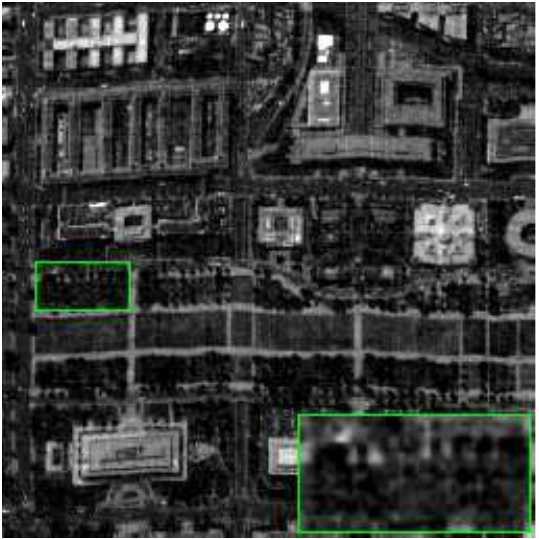} \\
    \includegraphics[width=\textwidth]{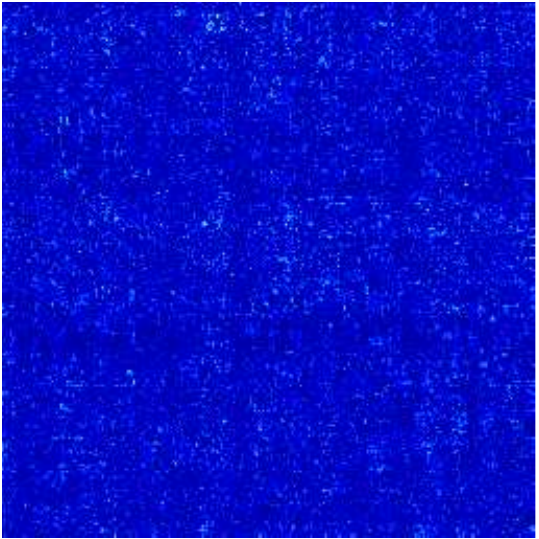} \\
    \end{minipage}
    \begin{minipage}[t]{0.09\textwidth}\centering
      \includegraphics[width=\textwidth]{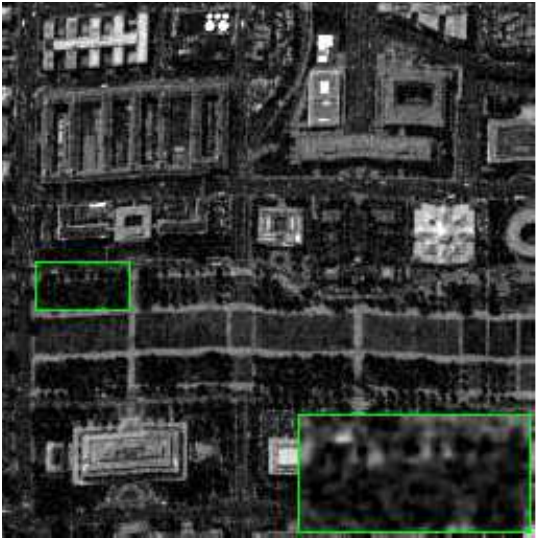} \\
    \includegraphics[width=\textwidth]{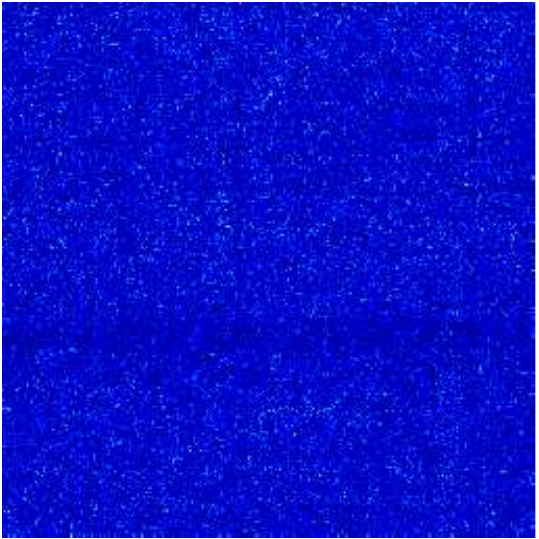} \\
    \end{minipage}
      \begin{minipage}[t]{0.09\textwidth}\centering
        \includegraphics[width=\textwidth]{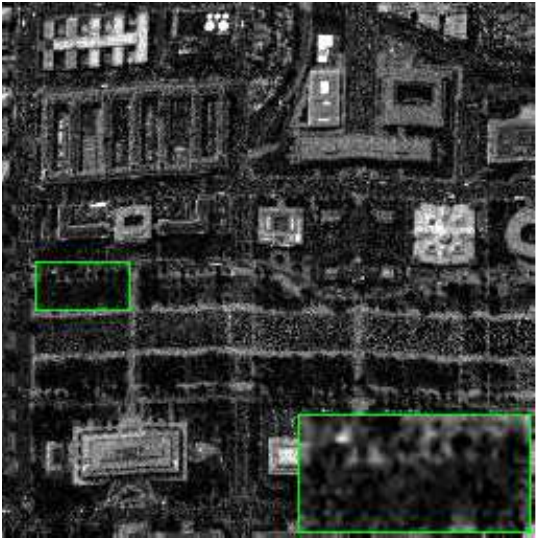}
    \includegraphics[width=\textwidth]{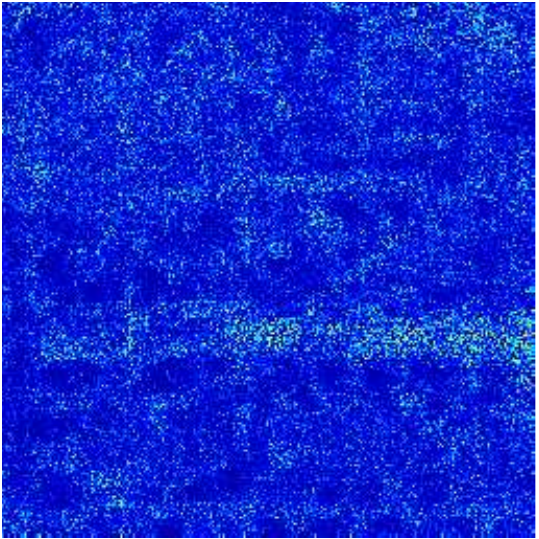} \\
    \end{minipage}
    \begin{minipage}[t]{0.09\textwidth}\centering
     \includegraphics[width=\textwidth]{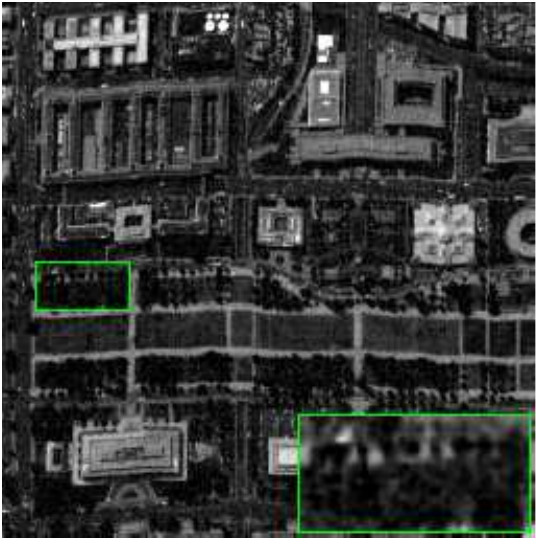} \\
  \includegraphics[width=\textwidth]{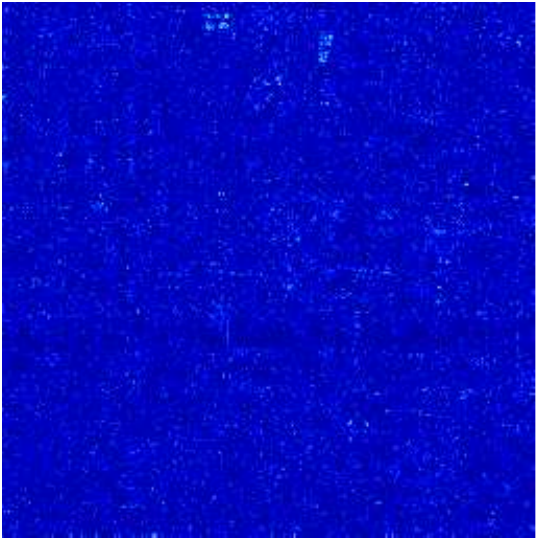} \\
  \end{minipage}
  \begin{minipage}[t]{0.09\textwidth}\centering
   \includegraphics[width=\textwidth]{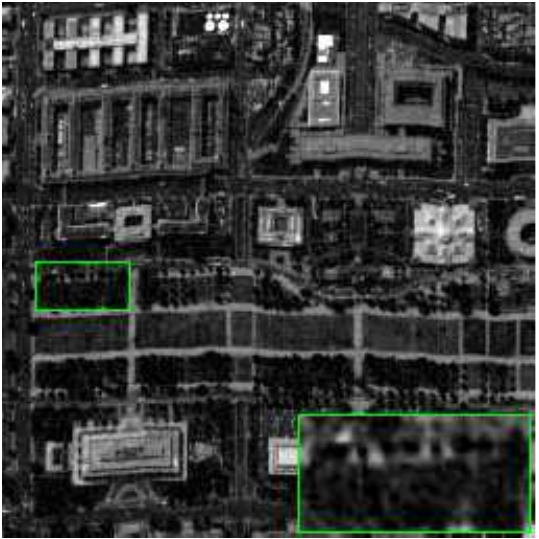} \\
  \includegraphics[width=\textwidth]{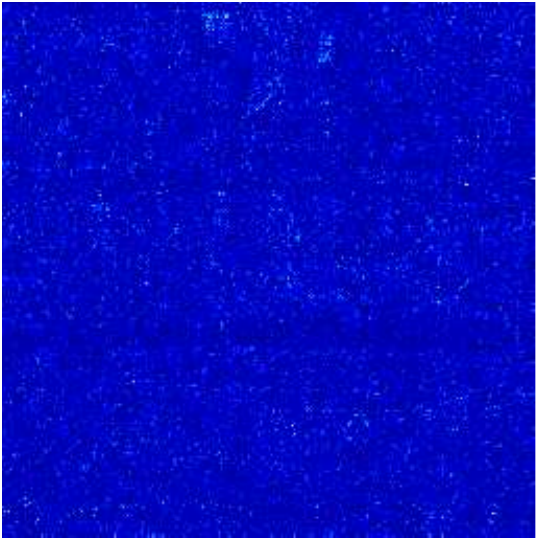} \\
  \end{minipage}
  \begin{minipage}[t]{0.09\textwidth}\centering
   \includegraphics[width=\textwidth]{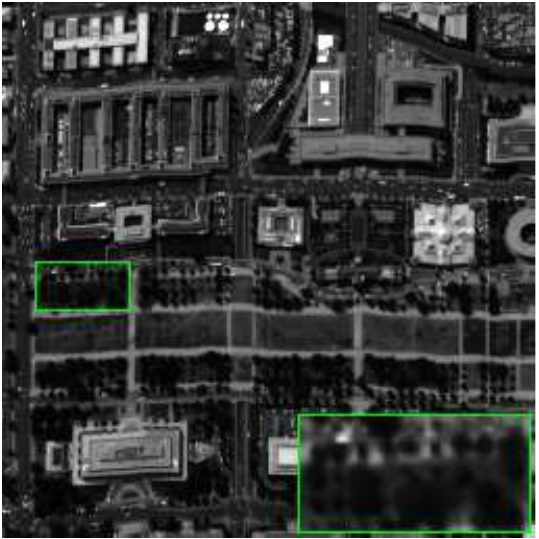} \\
  \includegraphics[width=\textwidth]{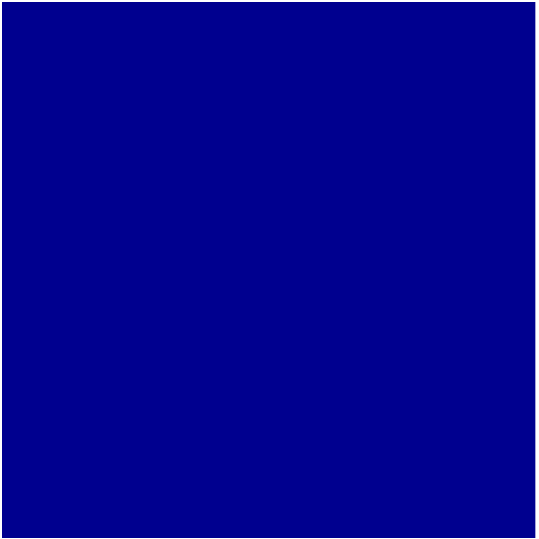} \\
  \end{minipage}
     \begin{minipage}{0.03\textwidth}\centering
  \includegraphics[width=0.95\textwidth,height=5.8\textwidth]{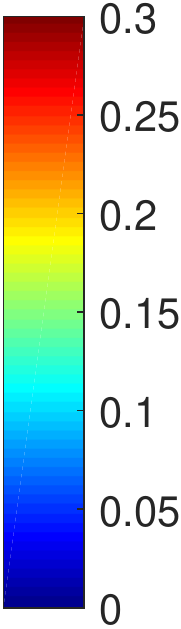} \\
  \end{minipage} \\
 \begin{minipage}[t]{0.09\textwidth}\centering
     \includegraphics[width=\textwidth]{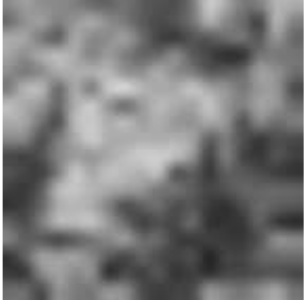} \\
    \includegraphics[width=\textwidth]{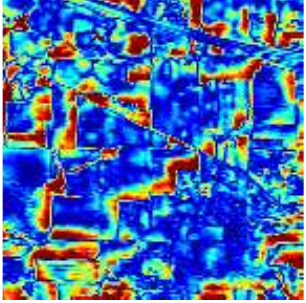} \\ \footnotesize (a) HSI
    \end{minipage}
    \begin{minipage}[t]{0.09\textwidth}\centering
     \includegraphics[width=\textwidth]{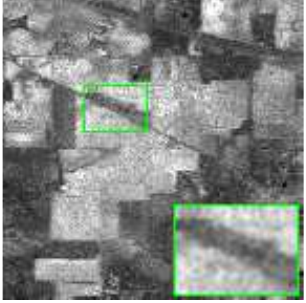} \\
    \includegraphics[width=\textwidth]{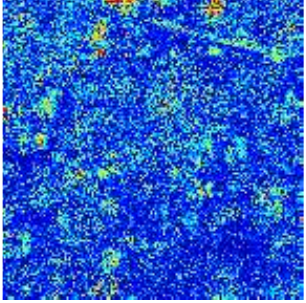} \\ \footnotesize(b) CNMF
    \end{minipage}
    \begin{minipage}[t]{0.09\textwidth}\centering
     \includegraphics[width=\textwidth]{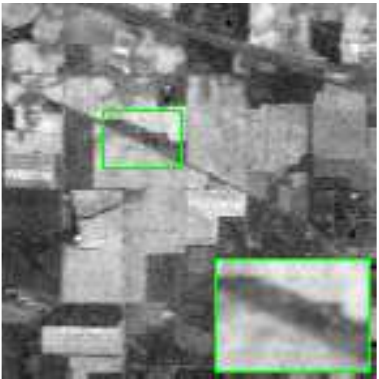} \\
    \includegraphics[width=\textwidth]{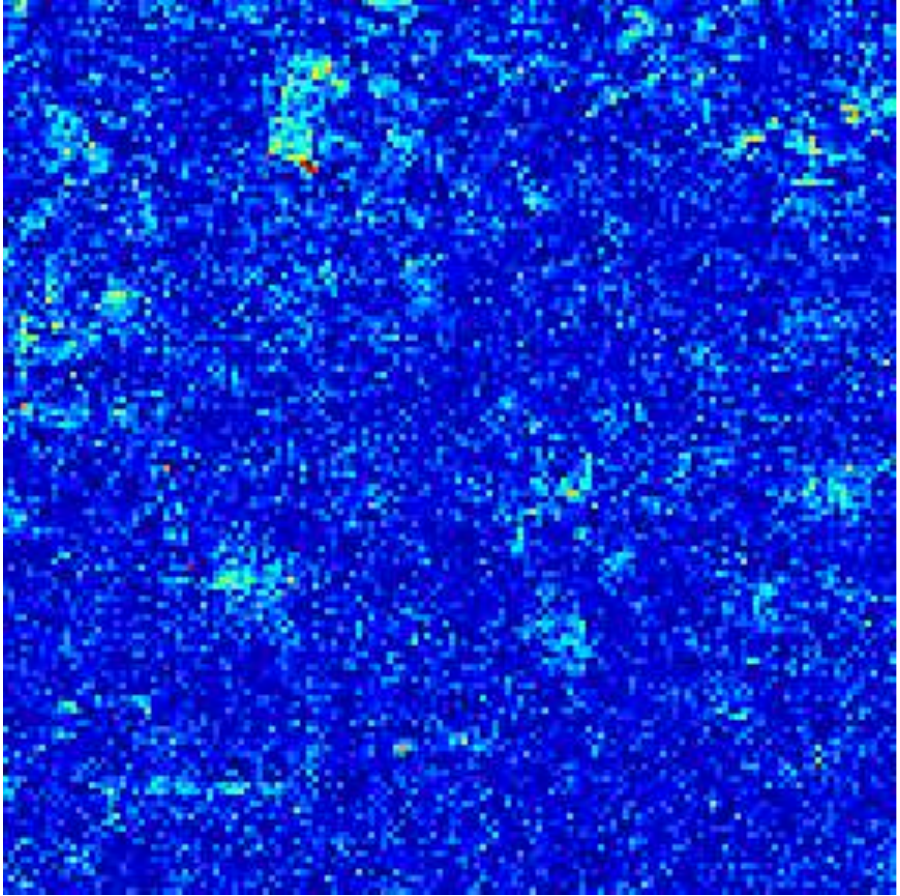} \\ \footnotesize(c) FUSE
    \end{minipage}
    \begin{minipage}[t]{0.09\textwidth}\centering
      \includegraphics[width=\textwidth]{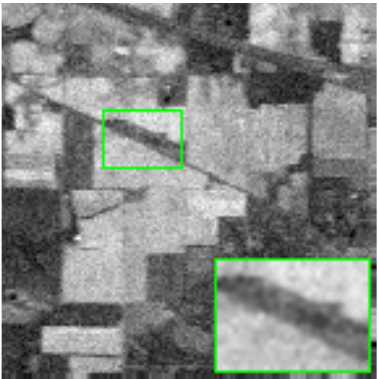} \\
    \includegraphics[width=\textwidth]{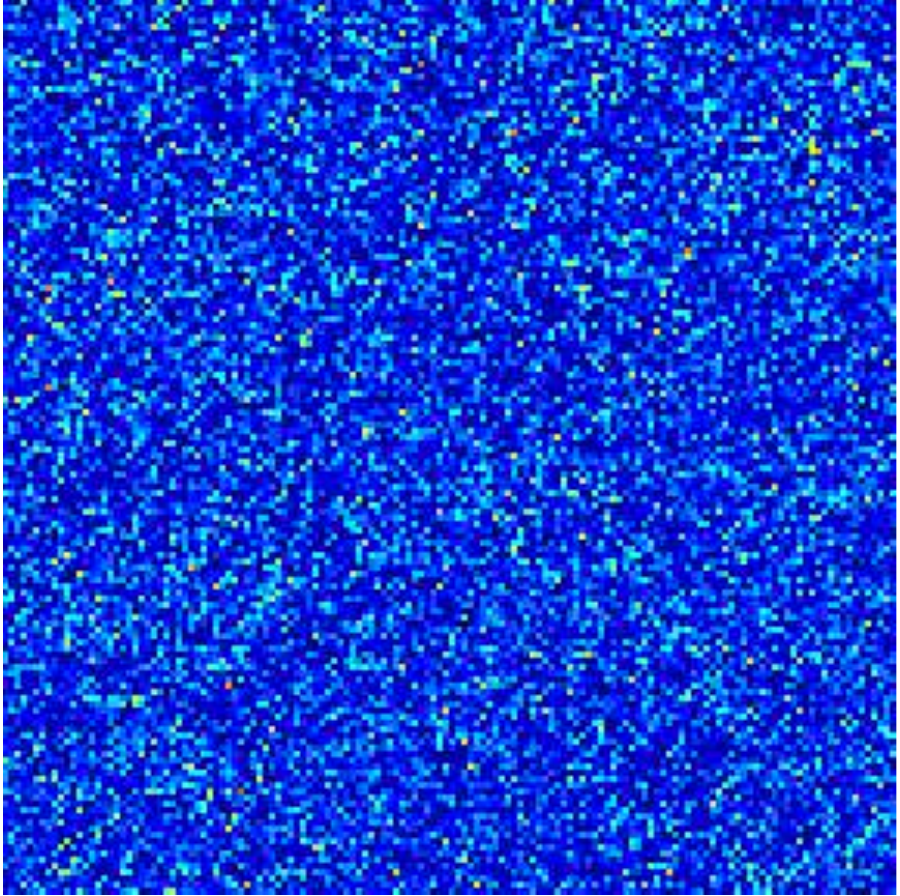} \\ \footnotesize(d) HySure
    \end{minipage}
    \begin{minipage}[t]{0.09\textwidth}\centering
        \includegraphics[width=\textwidth]{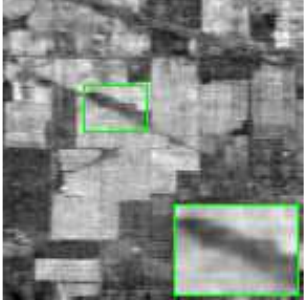} \\
    \includegraphics[width=\textwidth]{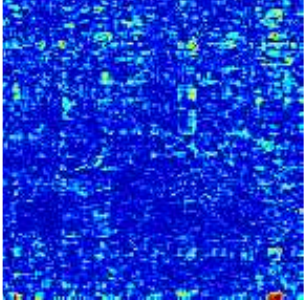} \\ \footnotesize(e) STEREO
    \end{minipage}
    \begin{minipage}[t]{0.09\textwidth}\centering
       \includegraphics[width=\textwidth]{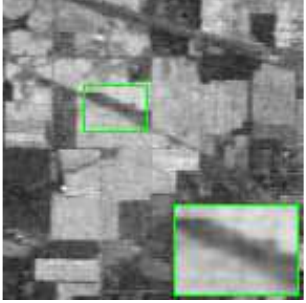} \\
    \includegraphics[width=\textwidth]{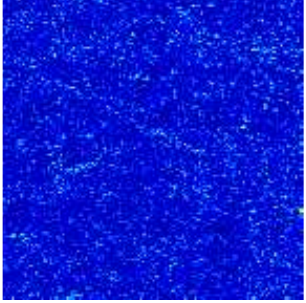} \\ \footnotesize(f) CSTF
    \end{minipage}
      \begin{minipage}[t]{0.09\textwidth}\centering
         \includegraphics[width=\textwidth]{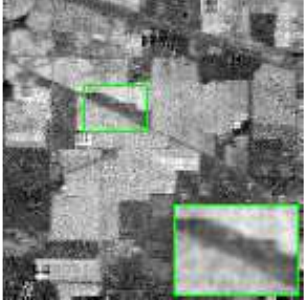}
    \includegraphics[width=\textwidth]{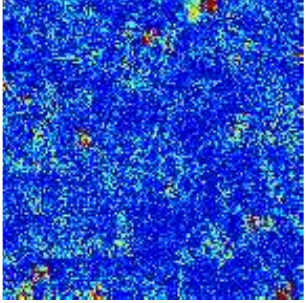} \\\footnotesize (g) NLSTF
    \end{minipage}
    \begin{minipage}[t]{0.09\textwidth}\centering
     \includegraphics[width=\textwidth]{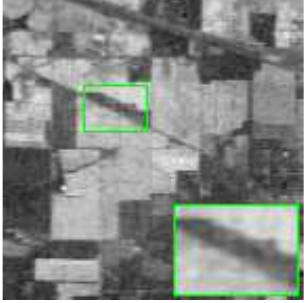} \\
  \includegraphics[width=\textwidth]{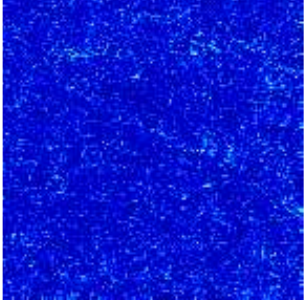} \\ \footnotesize(h) CTRF
  \end{minipage}
  \begin{minipage}[t]{0.09\textwidth}\centering
    \includegraphics[width=\textwidth]{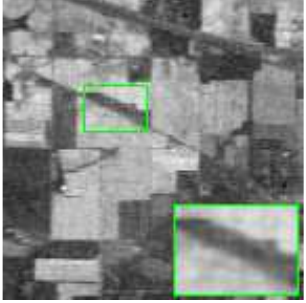} \\
  \includegraphics[width=\textwidth]{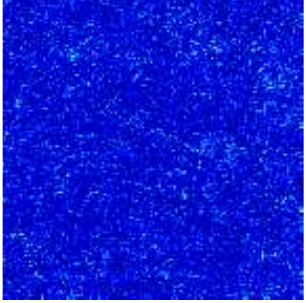} \\\footnotesize (i) NCTRF
  \end{minipage}
  \begin{minipage}[t]{0.09\textwidth}\centering
    \includegraphics[width=\textwidth]{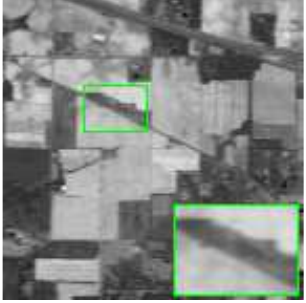} \\
  \includegraphics[width=\textwidth]{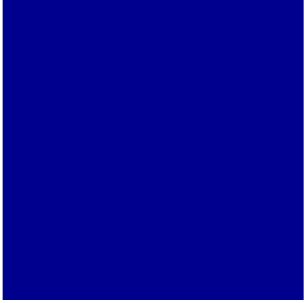} \\ \footnotesize (j) HR-HSI
  \end{minipage}
    \begin{minipage}{0.03\textwidth}\centering
  \includegraphics[width=\textwidth,height=5.8\textwidth]{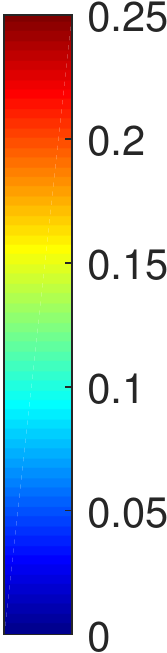} \\
  \end{minipage} \\
\caption{HSR results of different methods with WDC and Indian dataset. The noise level is $SNR=20$. The first row illustrates the band 13 images from WDC; the second row is the related difference images between HSR results and the ground truth HR-HSI. The third row illustrates the band 49 images from Indian, and the last row is the related difference images.}
\vspace{-7px}
\label{fig:reconstruct}
\end{figure*}
\subsection{Comparison methods and evaluation measures}
We choose the following methods for comparison: coupled non-negative matrix factorization (CNMF)~\cite{yokoya2012coupled}
\footnote{\url{http://naotoyokoya.com/Download.html}},
FUSE~\cite{Qi_BFUSE}
\footnote{\url{https://github.com/qw245/BlindFuse}},
HySure~\cite{simoes2015convex}
\footnote{\url{https://github.com/alfaiate}},
coupled CP factorization (STEREO)~\cite{kanatsoulis2018hyperspectral}
\footnote{\url{https://sites.google.com/site/harikanats/}},
coupled Tucker factorization (CSTF)~\cite{li2018fusing}
\footnote{\url{https://sites.google.com/view/renweidian/}},
and
non-local sparse tensor factorization (NLSTF)~\cite{dian2017hyperspectral}.
The related implementation codes are downloaded from the authors' website and the parameters are manually tuned to the best. Our proposed methods are denoted as CTRF and NCTRF. \par
To evaluate the performance of the proposed method, five quantitative indices are utilized in our study, including peak signal-to-noise ratio (PSNR), root mean square error (RMSE), relative dimensional
global error in synthesis (ERGAS) \cite{wald2000quality}, spectral angle mapper (SAM), structure similarity (SSIM) \cite{wang2004image}. The smaller RMSE, ERGAS, and SAM indicate better super-resolution results. On the contrary, the larger PSNR and SSIM illustrate the better quality.

\subsection{Experimental results}

\noindent
\textbf{Quantitative  comparison.}
For each case of noise level, we calculate the evaluation values of three datasets and then average them, as presented in Table~\ref{tab:QEva}. From the table, it can be observed that the proposed CTRF and NCTRF have more advantages in the noise case of SNR equal to 20dB and 30dB. In the low-noise case, NLSTF achieves the best accuracy. This is mainly because the non-local based methods are more effective compared to global based methods. However, the increase of noise will bring a huge challenge to the group matching model, resulting in the performance decrease of NLSTF in the noisy case. From the comparison, the CNMF and HySure perform the worst, indicating the advantage of the coupled tensor model. Furthermore, CTRF and NCTRF achieve better results than those of STEREO and CSTF, demonstrating the advantage of TR decomposition compared to Tucker and CP decomposition. Finally, the results of NCTRF is slightly better than that of CTRF. This phenomenon demonstrates the advantage of nuclear norm regularization to the third core tensor. As the increase of noise level, the gap between CTRF and NCTRF decreases, due to the fact that in the noisy model, we usually choose a smaller TR \textit{rank} $R=[R_1,R_2,R_3]$. In this case, the efficiency of nuclear norm regularization also decreases, since the value $R_1 \times R_3$ now is small enough to explore the global spectral low-rank property of HR-HSI.
\begin{figure*}[!t]
  \centering
  \begin{minipage}[t]{0.245\textwidth}\centering
    \includegraphics[width=\textwidth]{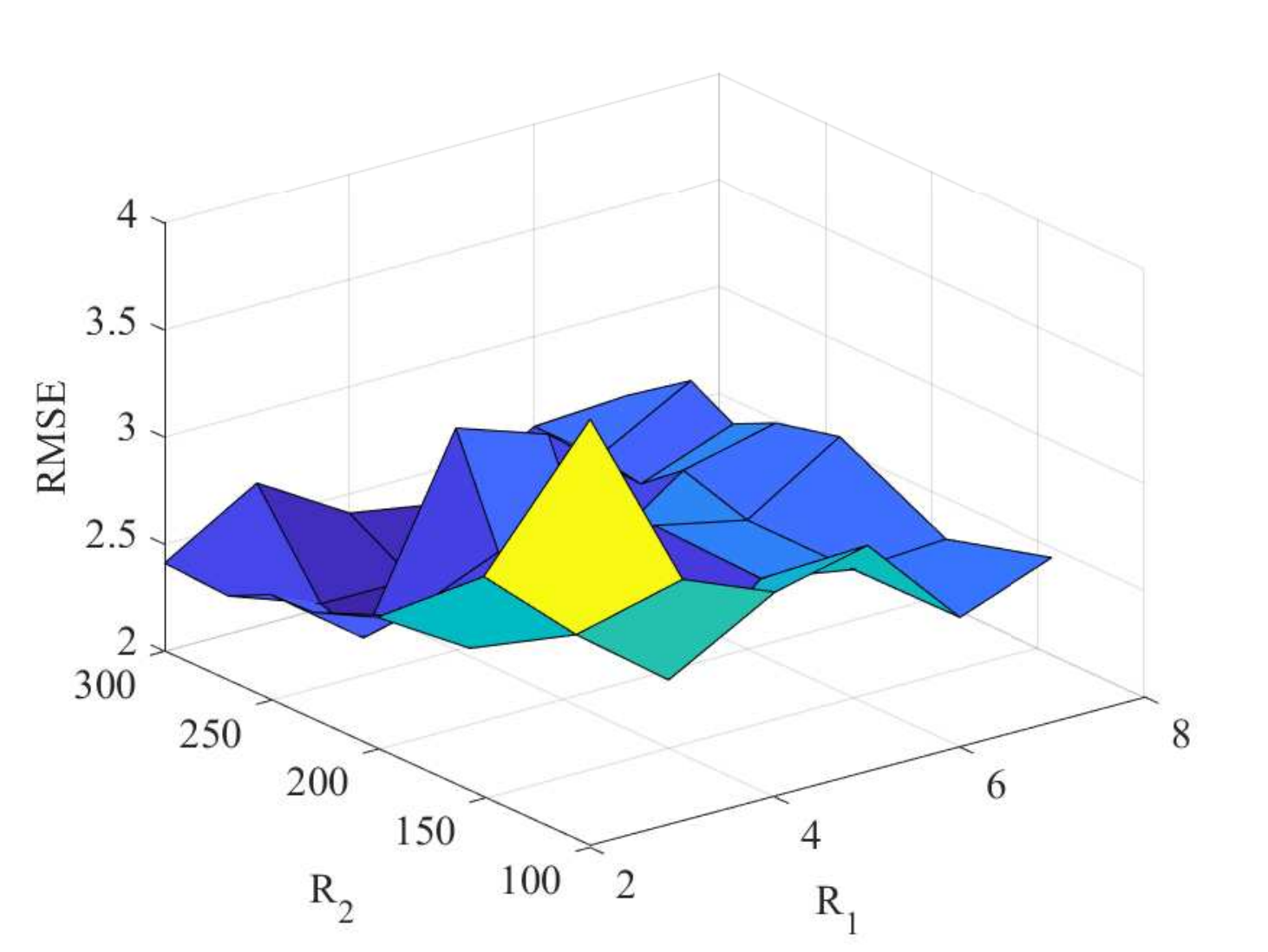} \\ \footnotesize (a) CTRF-WDC
    \end{minipage}
    \begin{minipage}[t]{0.245\textwidth}\centering
    \includegraphics[width=\textwidth]{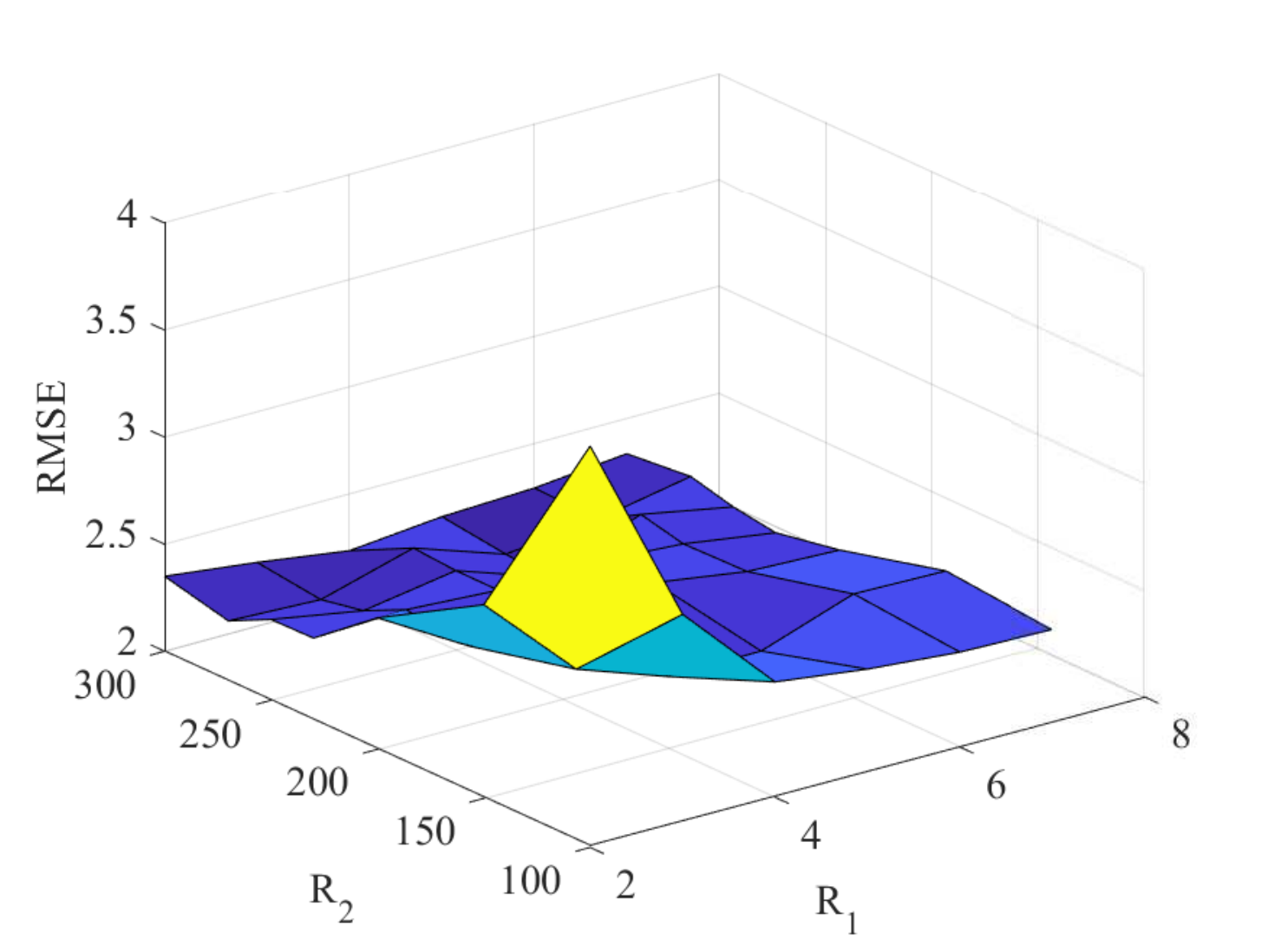} \\\footnotesize (b) NCTRF-WDC
    \end{minipage}
    \begin{minipage}[t]{0.245\textwidth}\centering
    \includegraphics[width=\textwidth]{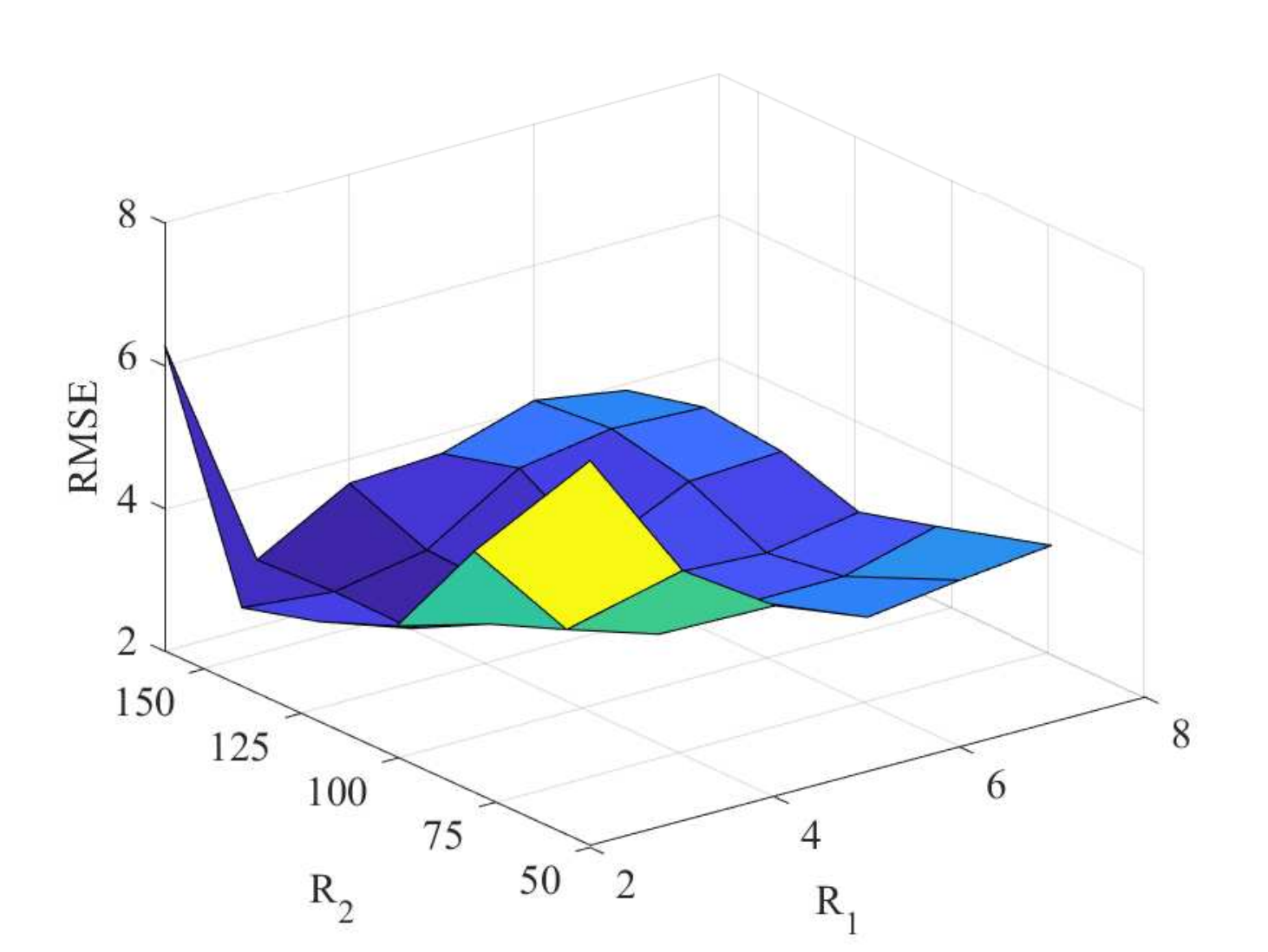} \\\footnotesize (c) CTRF-Indian
    \end{minipage}
    \begin{minipage}[t]{0.245\textwidth}\centering
    \includegraphics[width=\textwidth]{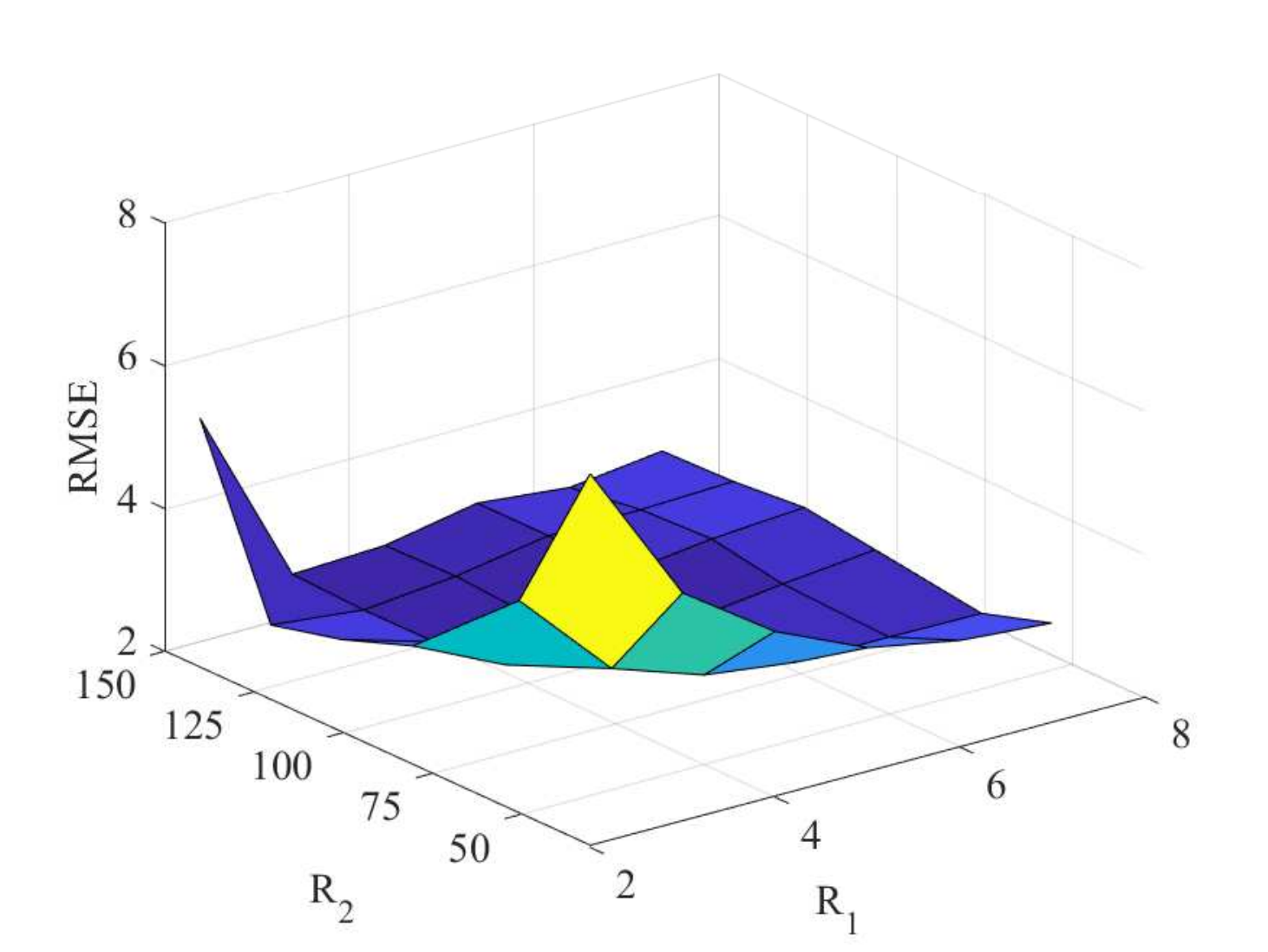} \\\footnotesize (d) NCTRF-Indian
    \end{minipage}
  \caption{The changes of RMSE value with different TR \textit{rank} [$R_1$, $R_2$, $R_1$]. The noise level is the case of SNR=40.}
\vspace{-7px}
\label{fig:Prank}
\end{figure*}

\begin{figure}[!t]
  \centering
  \begin{minipage}[t]{0.23\textwidth}\centering
  \includegraphics[width=\textwidth]{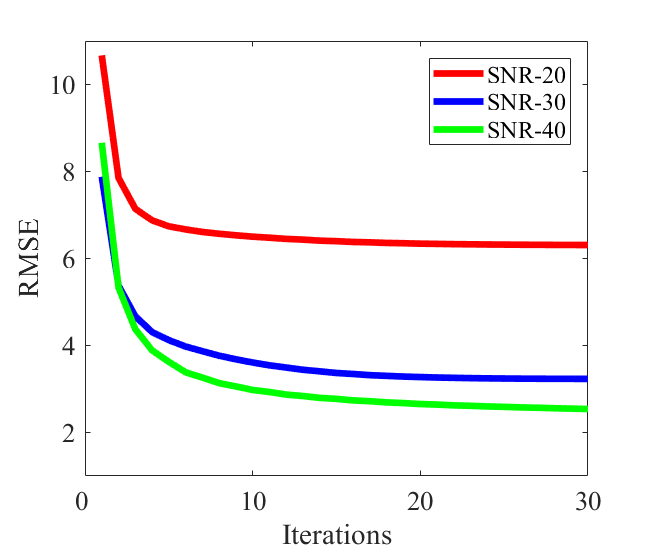} \\\footnotesize (a) CTRF
  \end{minipage}
  \begin{minipage}[t]{0.23\textwidth}\centering
  \includegraphics[width=\textwidth]{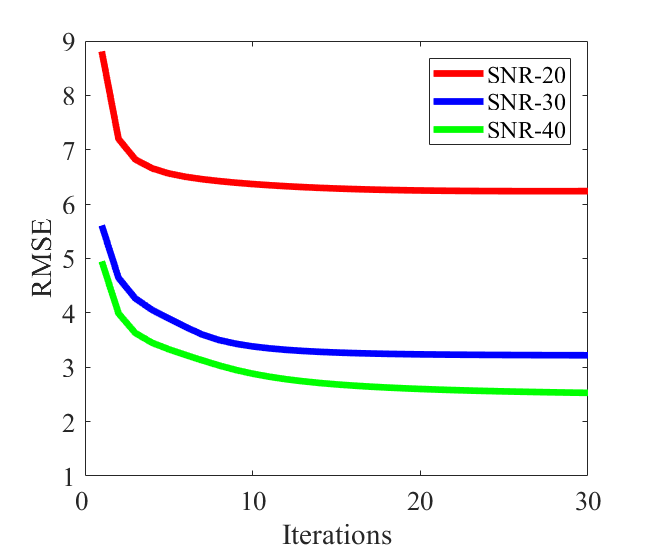}
  \\ \footnotesize (b) NCTRF
  \end{minipage} \\
\caption{The changes of RMSE value with the iterations on the WDC dataset.}
\vspace{-7pt}
\label{fig:convergence}
\end{figure}

\noindent
\textbf{Visual comparison.}
To further compare the differences of different HSR methods, we choose one band from WDC and one band from Indian Pines, to illustrate the related HSI, ground truth HR-HSI, and different HSR results in the case of noise level $SNR=20$ in Figure \ref{fig:reconstruct}. We also present the related difference images between the HSR results and the ground truth. From the figure, it can be observed that the proposed CTRF and NCTRF achieve the best visual results. CSTF can also achieve outstanding results. STEREO and FUSE perform better in the case of the WDC dataset, but worse in the Indian Pines case. CNMF, HySure, and NLSTF fail to reconstruct the image.

\subsection{Discussion}
\label{sec:discuss}
\noindent
\textbf{Parameter analysis.} The TR \textit{rank} $R=[R_1,R_2,R_3]$ is the most important parameter in the proposed CTRF and NCTRF. Until now, the adaptive selection of TR \textit{rank} is still a key problem. From Theorem~\ref{th:1}, the values $R_1\times R_2, R_2 \times R_3, R_3 \times R_1$ are bounded by $rank(\mat{X}_{<1>}), rank(\mat{X}_{<2>})$ and $rank(\mat{X}_{<3>})$, respectively. To simplify the complexity of parameter analysis, we choose $R_1 = R_3$, and change $R_1$ from the range $[2, 7]$; meanwhile, change $R_2$ from the range $[50, 300]$. Figure~\ref{fig:Prank} presents the changes of RMSE value with different TR \textit{rank} [$R_1$, $R_2$, $R_1$]. Typically, when the rank $R_1$ is larger, the performance of CTRF degrades significantly. This is mainly because the subspace dimensions of some classes may smaller than $R_1$. However, with the nuclear norm constraint of the third core tensor, NCTRF can obtain smaller RMSE values, compared to that of CTRF. We choose TR $rank$ as $[3,150,3], [4,200,4], [5,250,5]$ for $SNR=20, 30, 40$, respectively.

\noindent
\textbf{Convergence.} We adopt experiments to demonstrate the convergence behavior of the proposed CTRF and NCTRF. Figure~ \ref{fig:convergence} presents the changes of RMSE obtained by CTRF and NCTRF, with the increase of iteration number on the WDC dataset. It can be observed that, with the increase of iterations, the RMSE values obtained by CTRF and NCTRF decrease to the stable values, indicating the convergence of the proposed methods.

\noindent
\textbf{Computational time.} Table~\ref{tab:timel} presents the computational time of different methods with different datasets. From the table, it can be observed that the proposed methods are competitive compared to the other methods.
\setlength{\tabcolsep}{0.7mm}{
\begin{table}[!t]
  \centering
  \footnotesize
  \caption{Computational time (s) of different methods with different dataset}
    \begin{tabular}{ccccccccc}
    \toprule
     Data  & CNMF & FUSE & HySure & STEREO & CSTF  & NLSTF & CTRF  & NCTRF \\
    \midrule
     WDC    & 15.8 &3.9 &50.2  & 5.3 & 15.1 & 20.7  & 12.0 & 17.0 \\
     Pavia  & 15.4 &3.8 &46.1  & 5.4 & 14.8 & 21.9  & 11.7 & 17.1 \\
     Indian & 3.1  &1.6 &18.6  & 0.9 & 3.6  & 13.2  & 2.2  & 3.3 \\
     \hline
    \end{tabular}%
    \vspace{-7px}
  \label{tab:timel}%
\end{table}}%

\setlength{\tabcolsep}{2mm}{
\begin{table}[htbp]
  \centering
  \caption{Quantitative comparison of different algorithms on CAVE Balloons and Toy images.}
  \footnotesize
    \begin{tabular}{c|ccc|ccc}
    \hline
    \multirow{2}[4]{*}{Methods} & \multicolumn{3}{c|}{CAVE\_Balloons} & \multicolumn{3}{c}{CAVE\_Toy} \\
\cmidrule{2-7}          & PSNR  & RMSE   & SSIM   & PSNR   & RMSE   & SSIM \\
    \hline
    CNMF  & 37.69 & 3.37  & 0.984 & 34.61 & 4.92  & 0.957 \\
    \hline
    FUSE  & 33.24 & 5.60  & 0.891 & 32.52 & 6.08  & 0.926 \\
    \hline
    HySure & 32.27 & 6.35  & 0.886 & 29.61 & 8.63  & 0.914 \\
    \hline
    STEREO & 37.61 & 3.78  & 0.948 & 34.56 & 5.10  & 0.907 \\
    \hline
    CSTF  & 40.51 & 2.75  & 0.979 & 37.78 & 3.95  & 0.957 \\
    \hline
    NLSTF & 40.02 & 4.17  & 0.986 & 38.41 & 3.96  & 0.979 \\
    \hline
    uSDN  & 37.78 & 3.45  & 0.965 & 33.94 & 5.71  & 0.949 \\
    \hline
    MHF-net    & 40.49 & \textbf{2.47}  & \textbf{0.989} & 36.10 & 4.18  & \textbf{0.981} \\
    \hline
    CTRF  & 40.57 & 2.70  & 0.972 & 37.58 & 4.04  & 0.951 \\
    \hline
    NCTRF & \textbf{41.51} & 2.49  & 0.979 & \textbf{38.42} & \textbf{3.88}  & 0.944 \\
    \hline
    \end{tabular}%
  \label{tab:DL}%
\end{table}}%

\begin{figure}[!t]
  \centering
   \begin{minipage}[t]{0.08\textwidth}\centering
        \includegraphics[width=\textwidth]{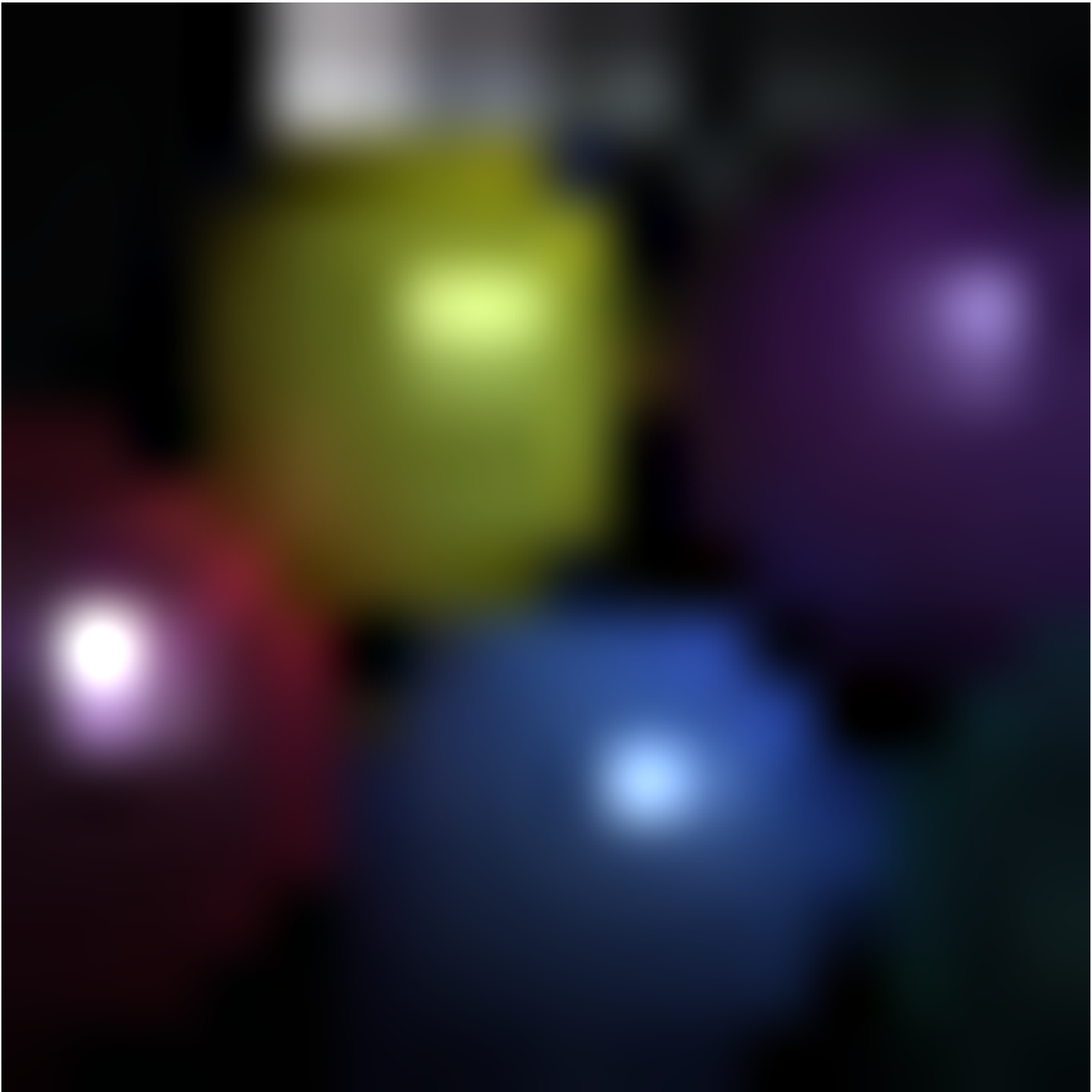} \\
    \includegraphics[width=\textwidth]{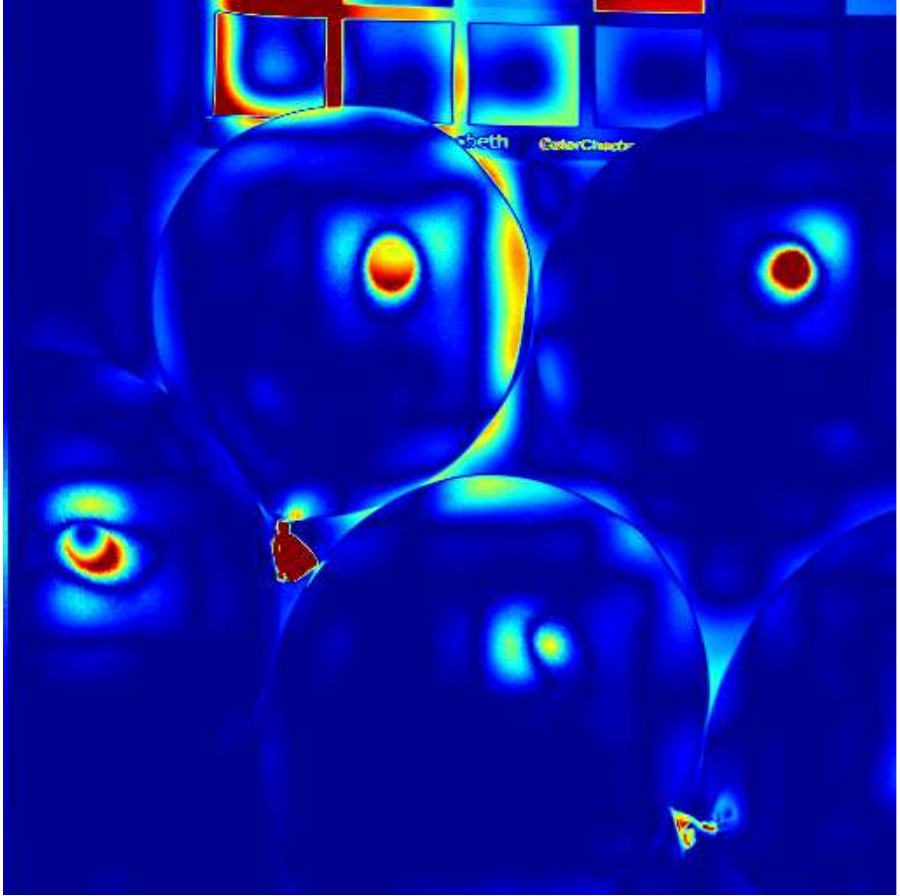} \\
    \end{minipage}
    \begin{minipage}[t]{0.08\textwidth}\centering
      \includegraphics[width=\textwidth]{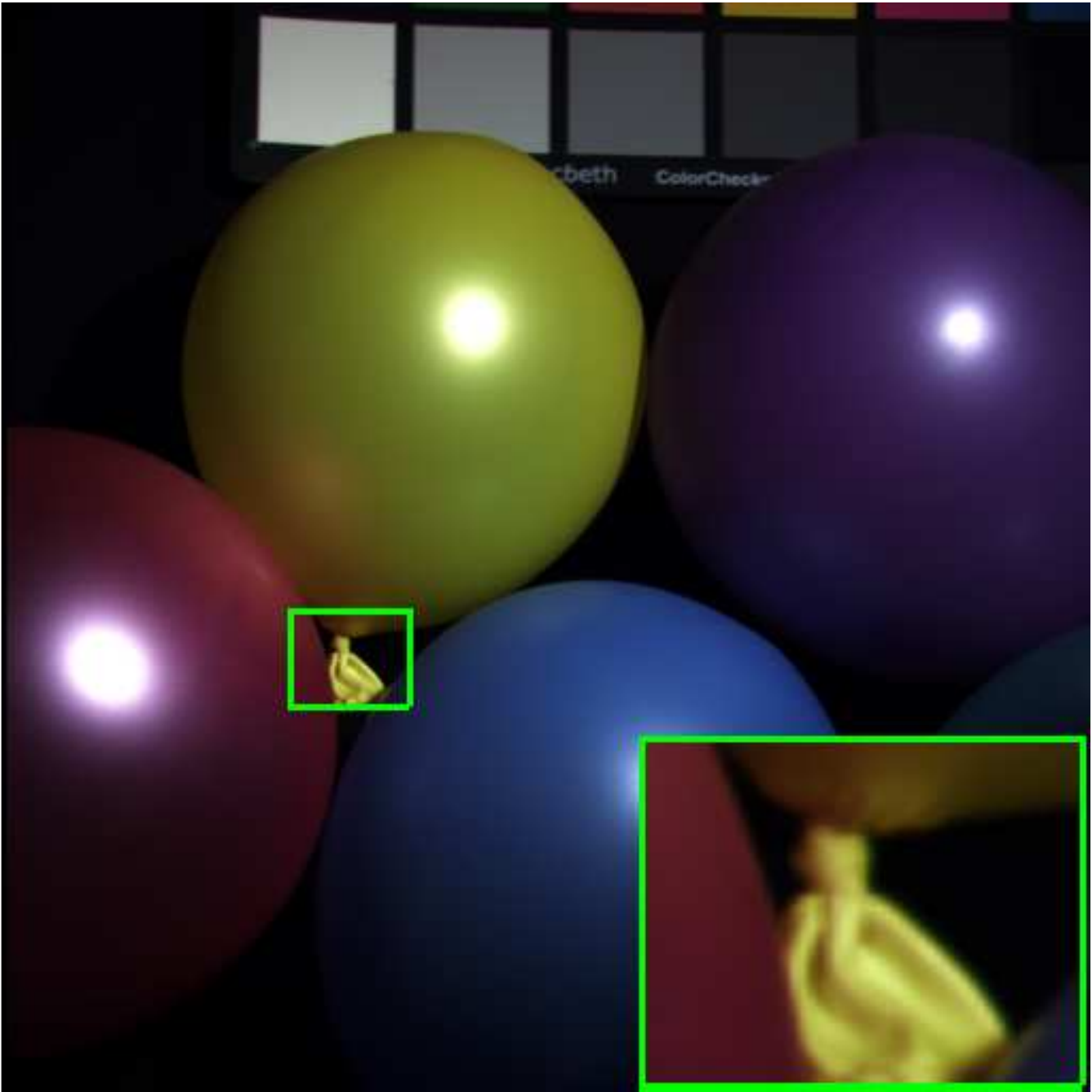} \\
    \includegraphics[width=\textwidth]{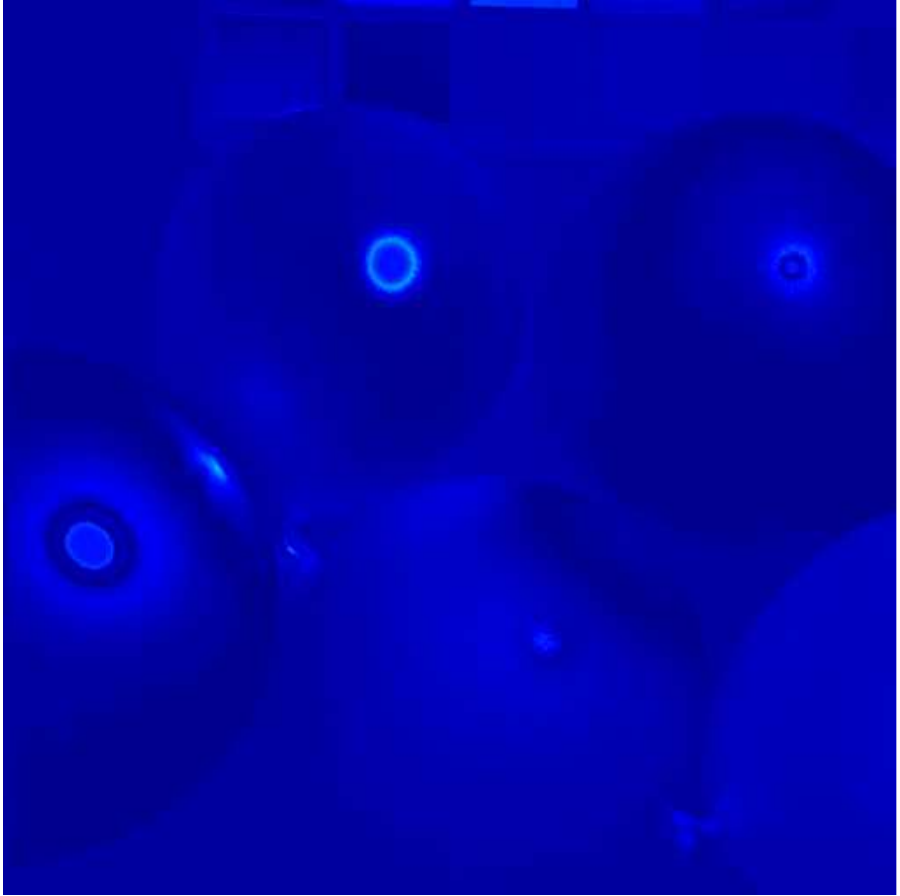} \\
    \end{minipage}
     \begin{minipage}[t]{0.08\textwidth}\centering
      \includegraphics[width=\textwidth]{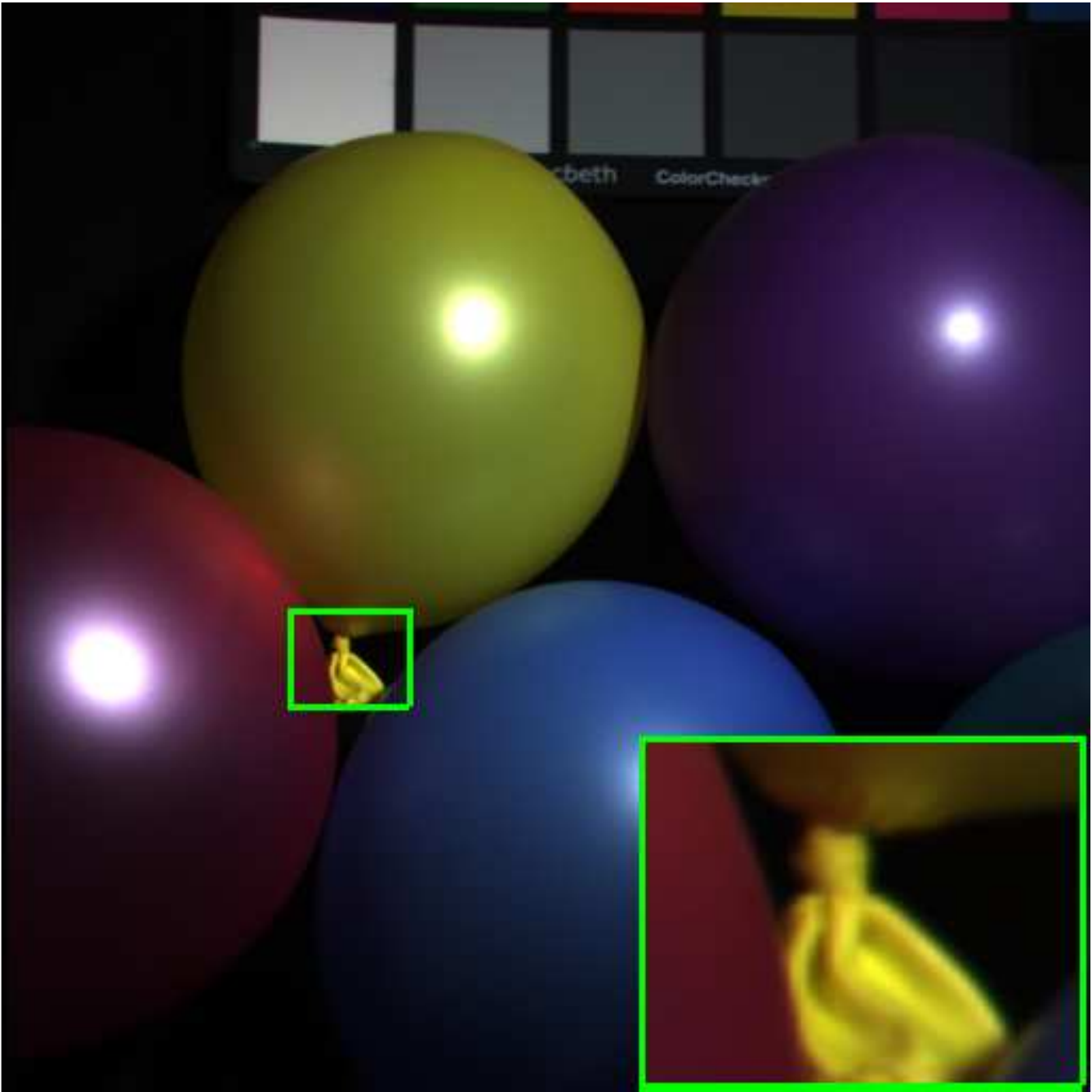} \\
    \includegraphics[width=\textwidth]{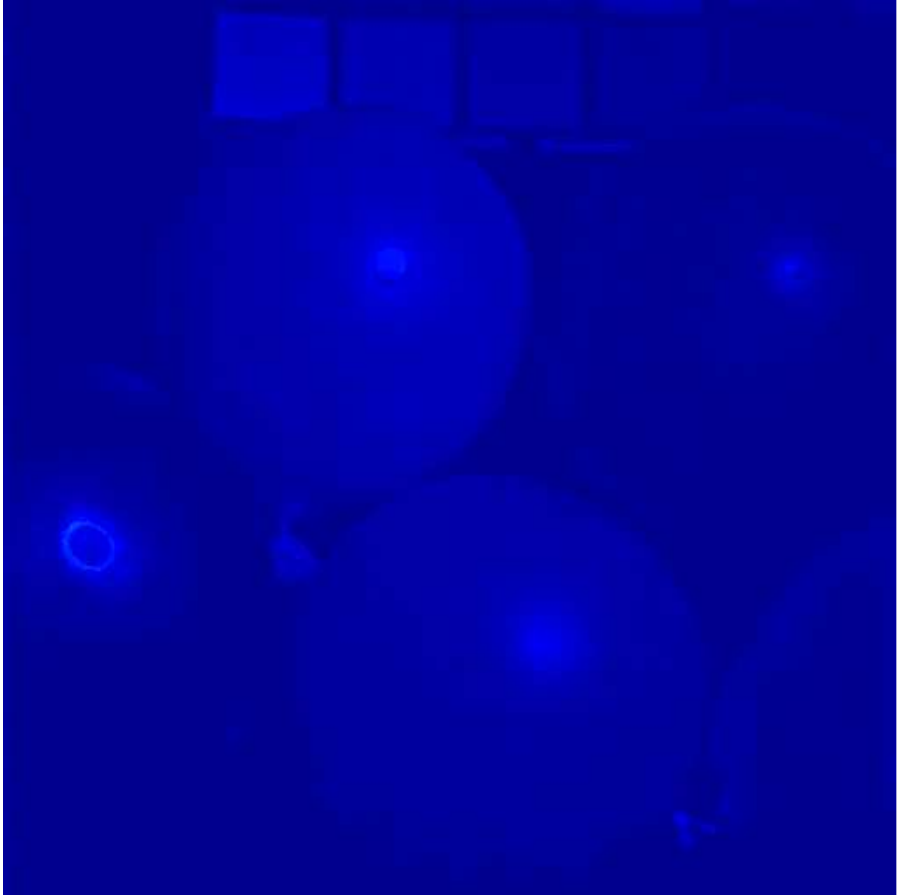} \\
    \end{minipage}
  \begin{minipage}[t]{0.08\textwidth}\centering
   \includegraphics[width=\textwidth]{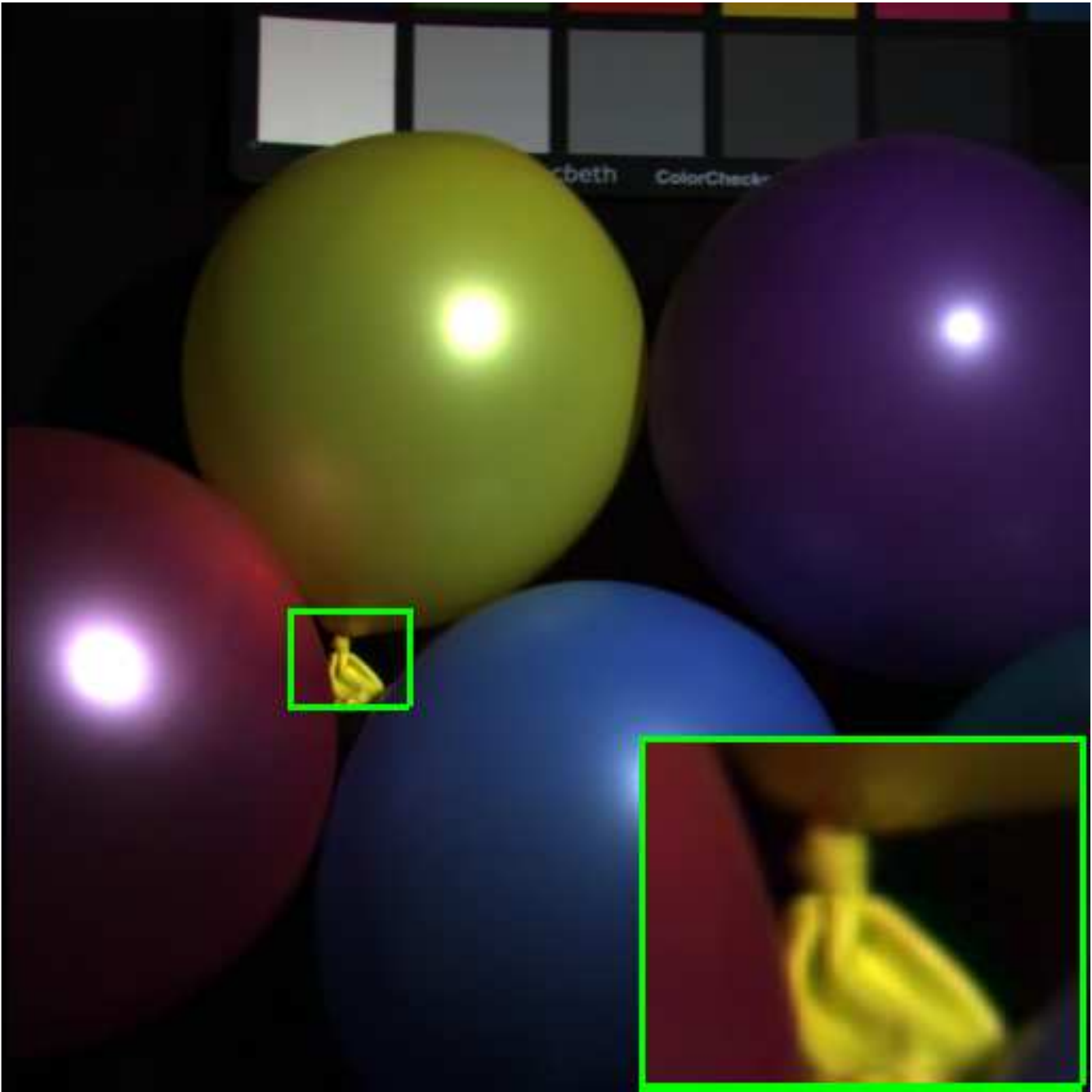} \\
    \includegraphics[width=\textwidth]{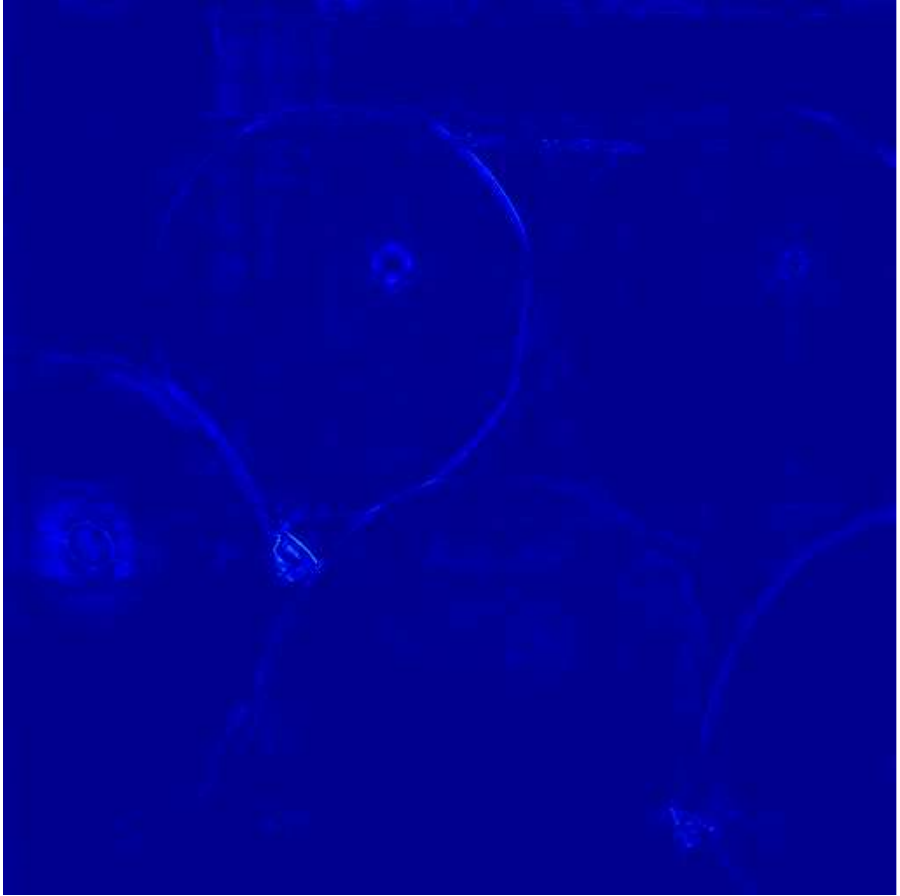} \\
    \end{minipage}
    \begin{minipage}[t]{0.08\textwidth}\centering
     \includegraphics[width=\textwidth]{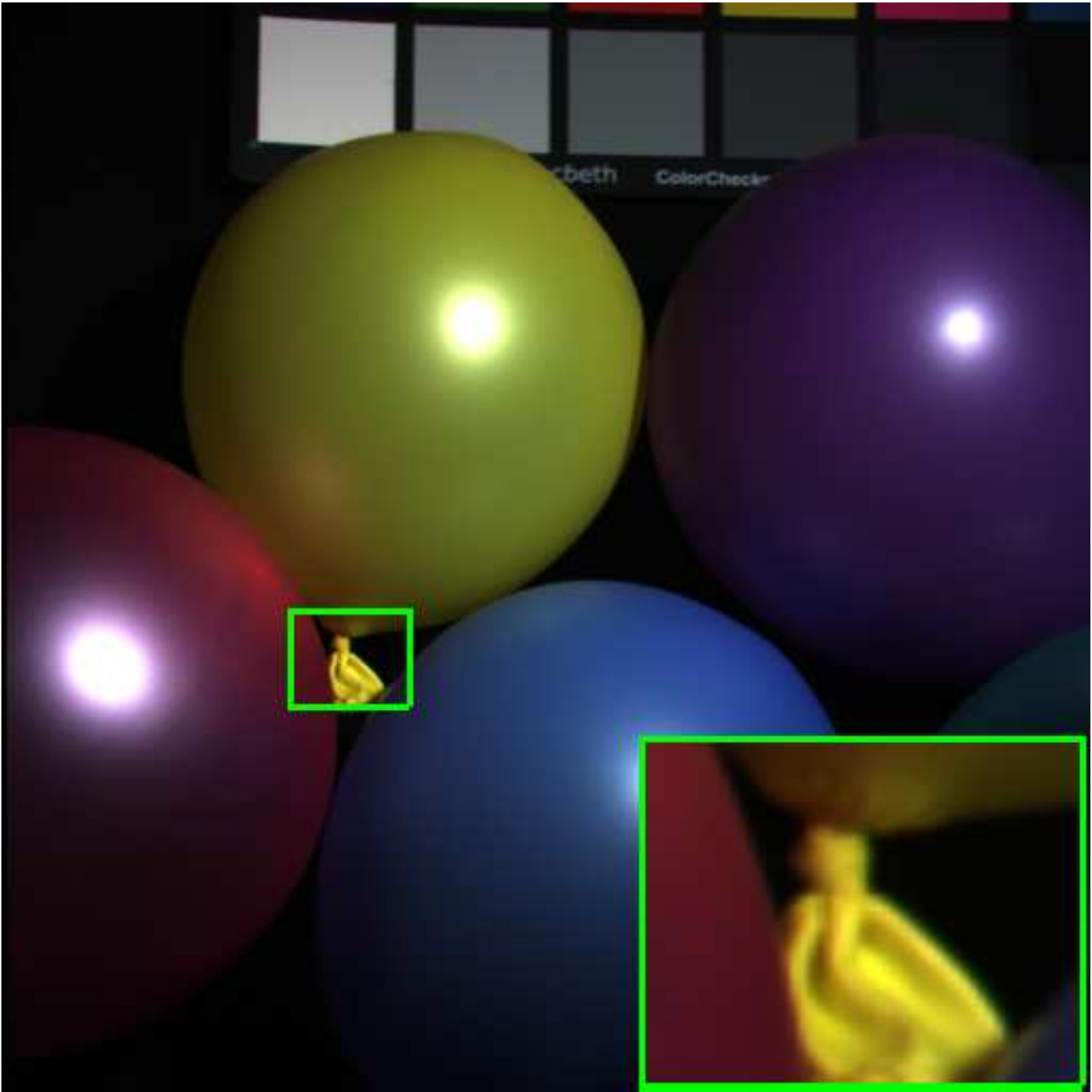} \\
    \includegraphics[width=\textwidth]{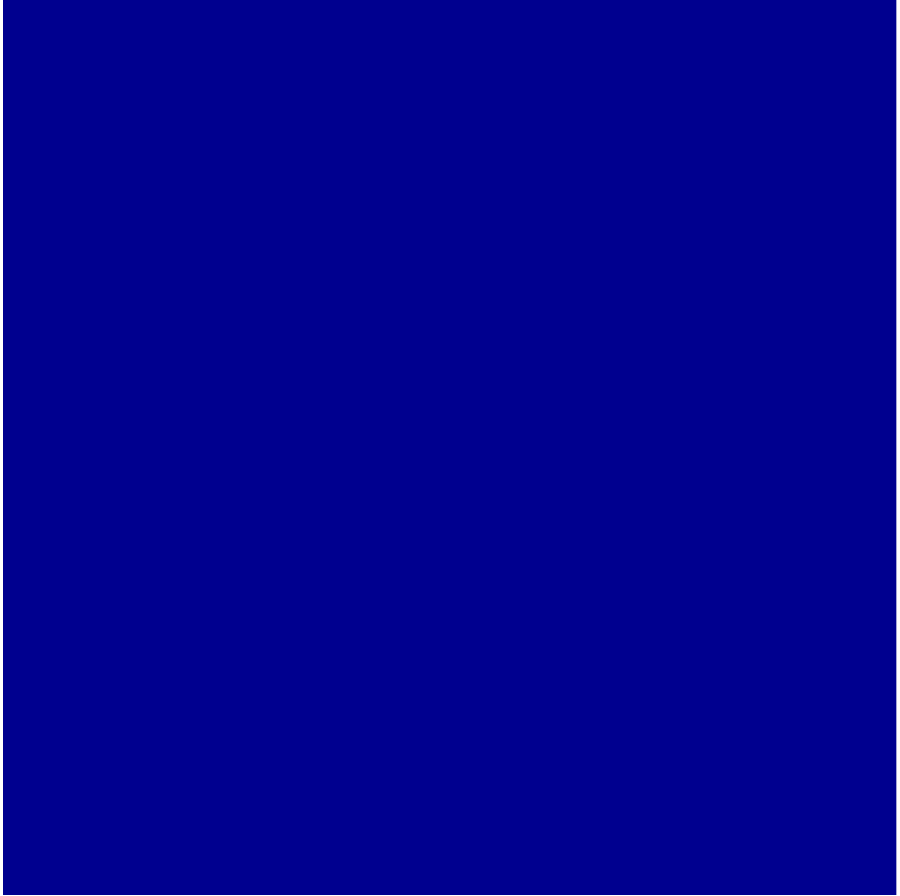} \\
    \end{minipage}
     \begin{minipage}{0.05\textwidth}\centering
  \includegraphics[width=0.95\textwidth,height=2.7\textwidth]{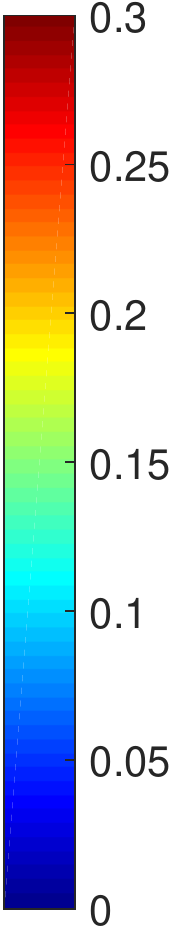}\centering \\
  \end{minipage} \\

    \begin{minipage}[t]{0.08\textwidth}\centering
        \includegraphics[width=\textwidth]{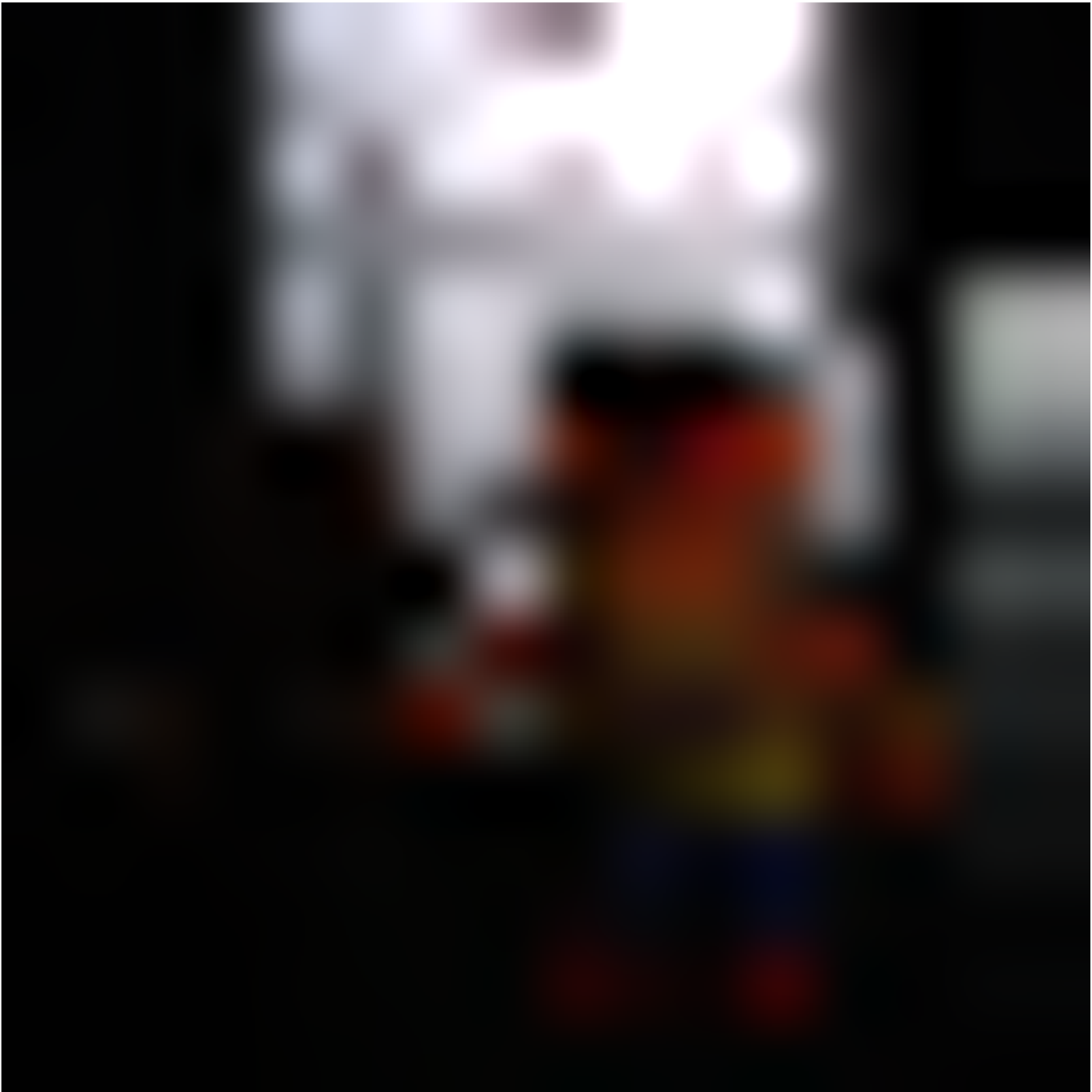} \\
    \includegraphics[width=\textwidth]{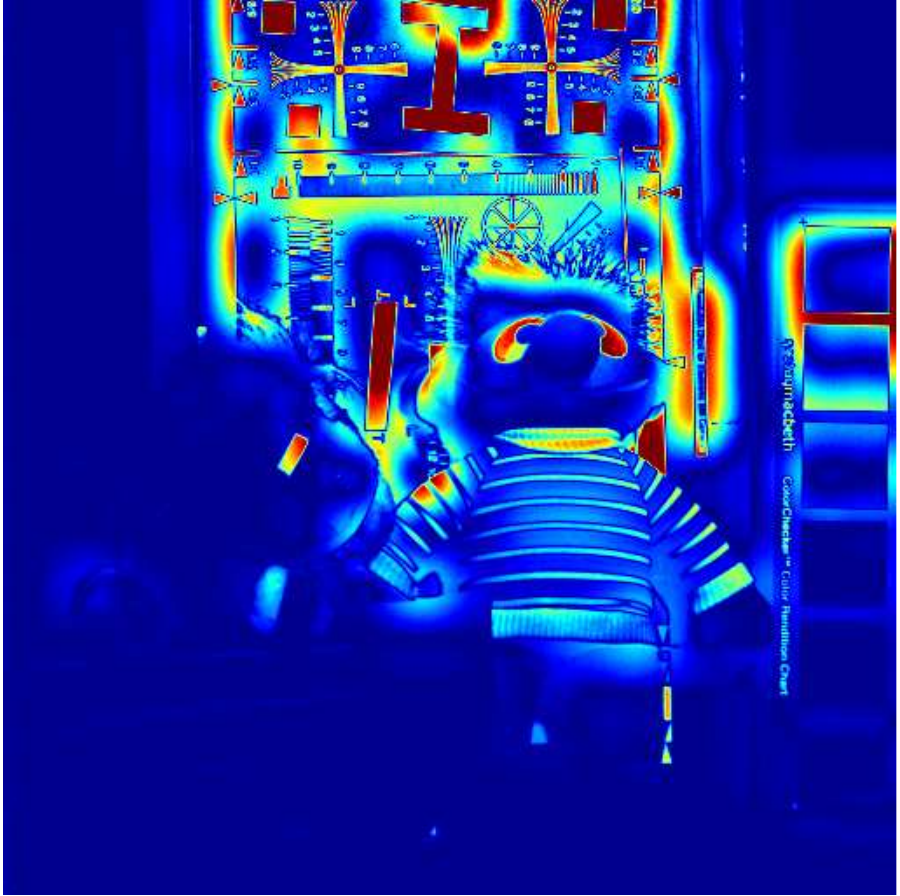} \\ \footnotesize(a) HSI
    \end{minipage}
    \begin{minipage}[t]{0.08\textwidth}\centering
      \includegraphics[width=\textwidth]{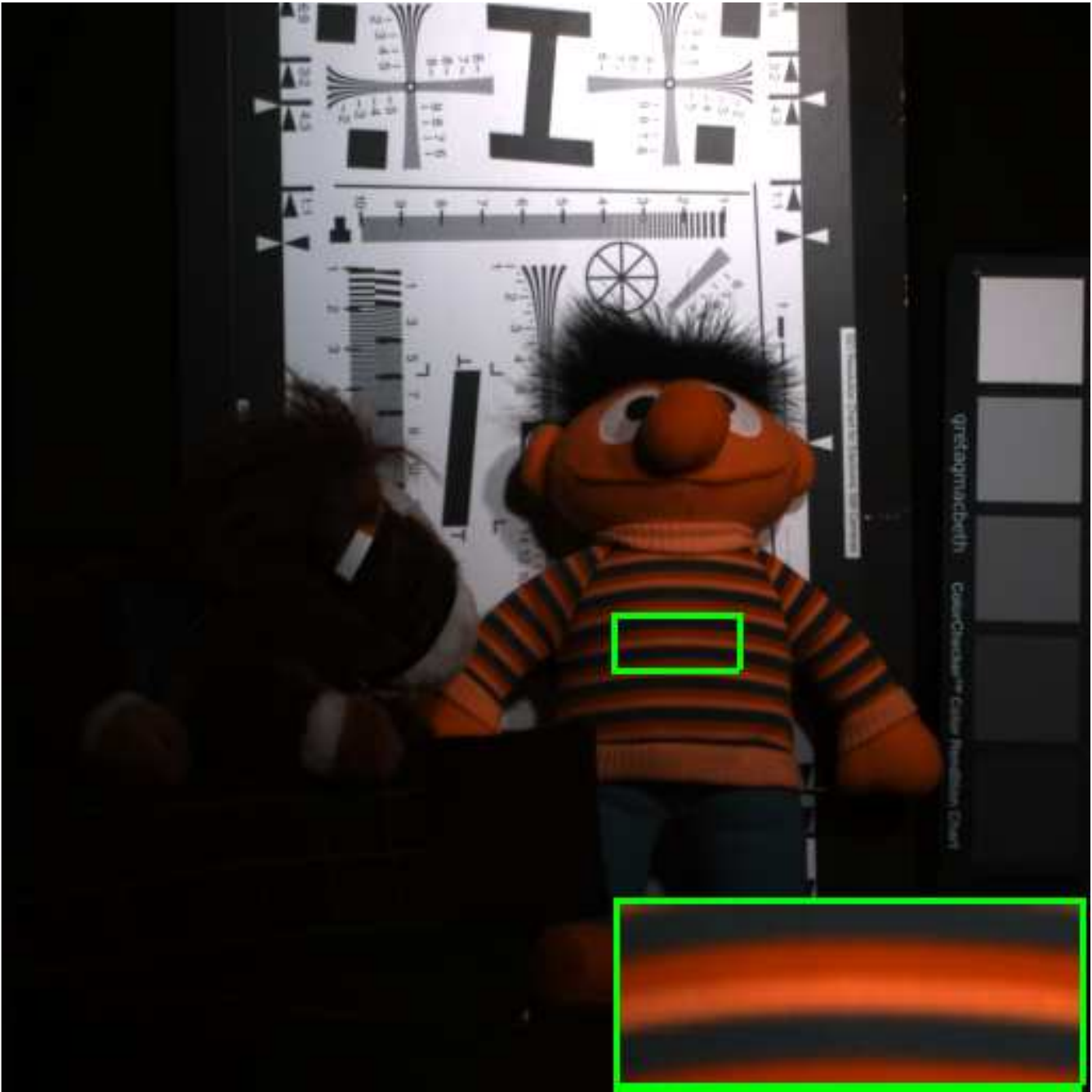} \\
    \includegraphics[width=\textwidth]{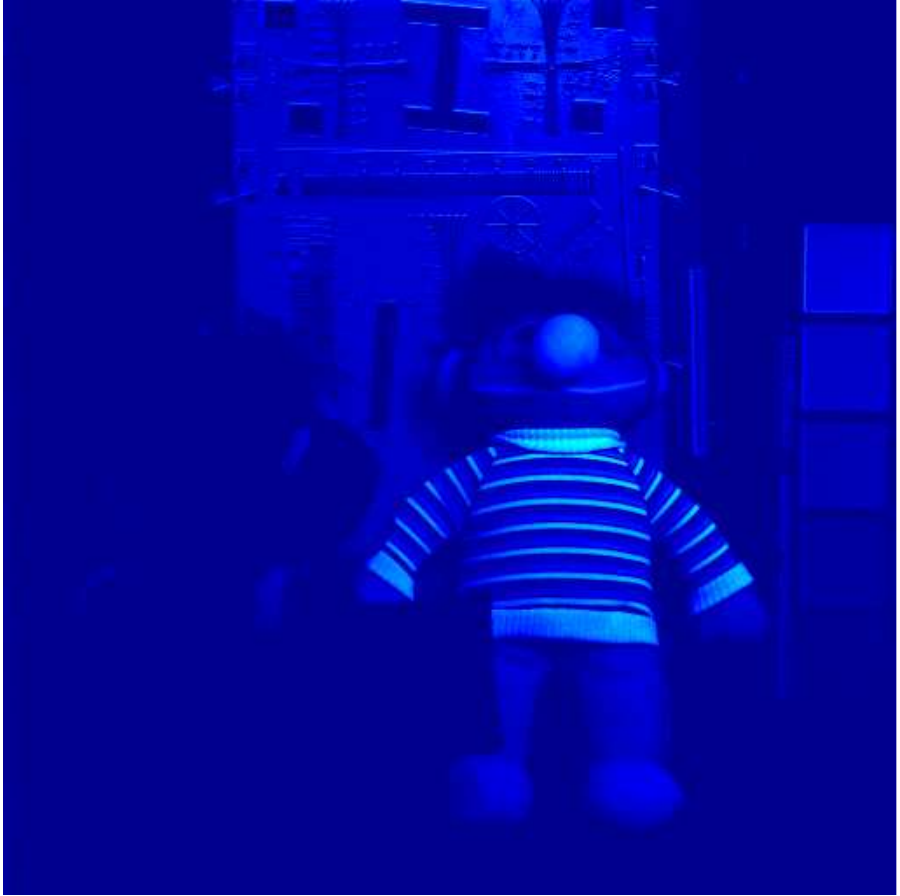} \\ \footnotesize(b) uSDN
    \end{minipage}
     \begin{minipage}[t]{0.08\textwidth}\centering
      \includegraphics[width=\textwidth]{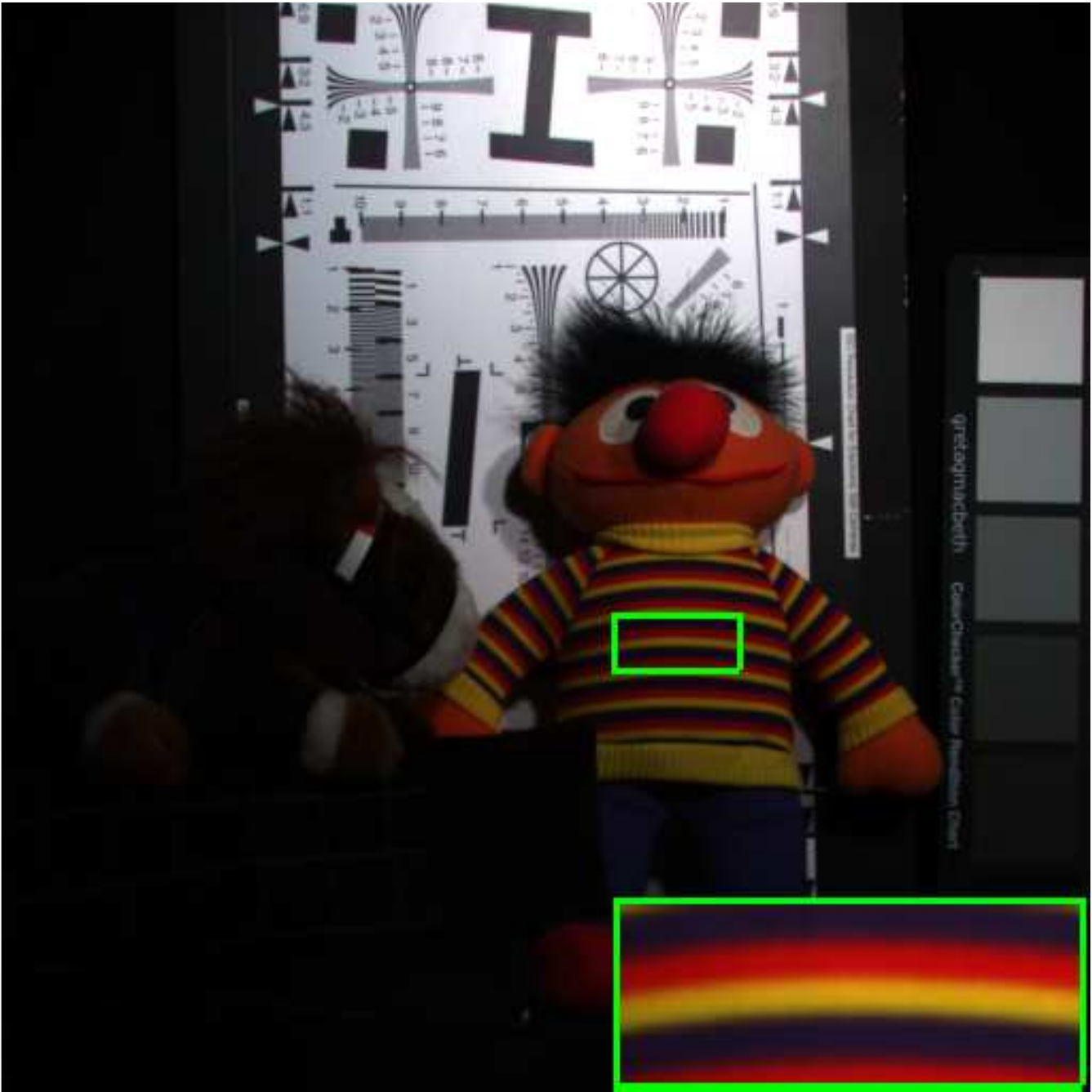} \\
    \includegraphics[width=\textwidth]{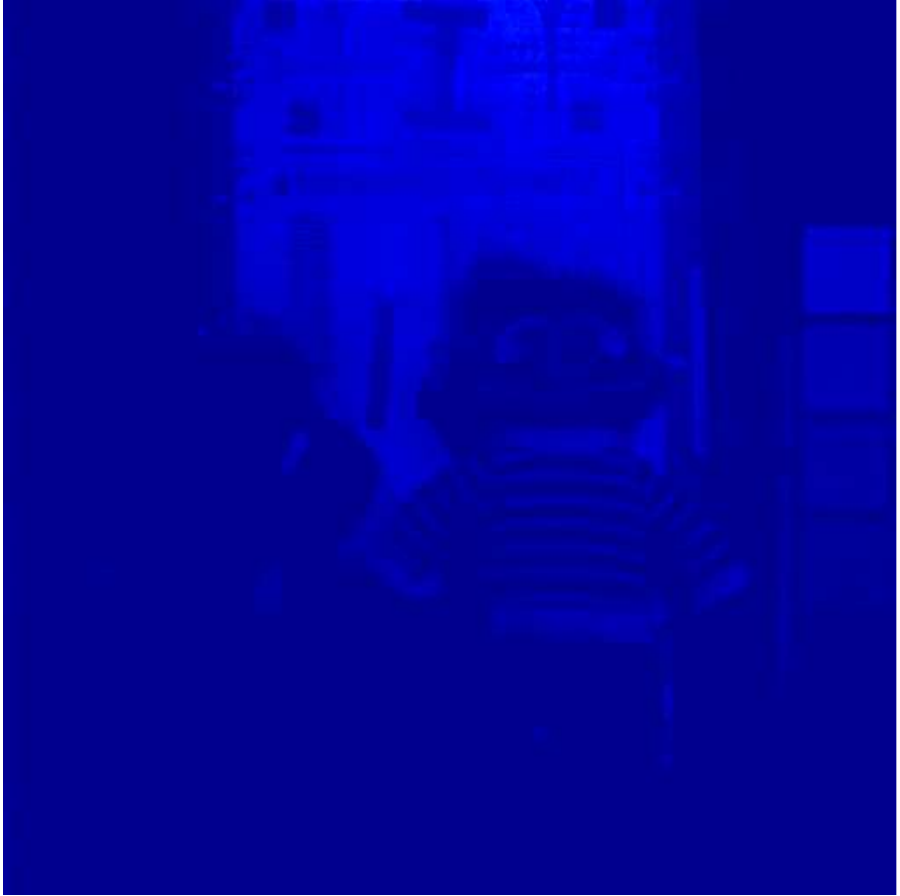} \\ \scriptsize(c) MHF-net
    \end{minipage}
  \begin{minipage}[t]{0.08\textwidth}\centering
   \includegraphics[width=\textwidth]{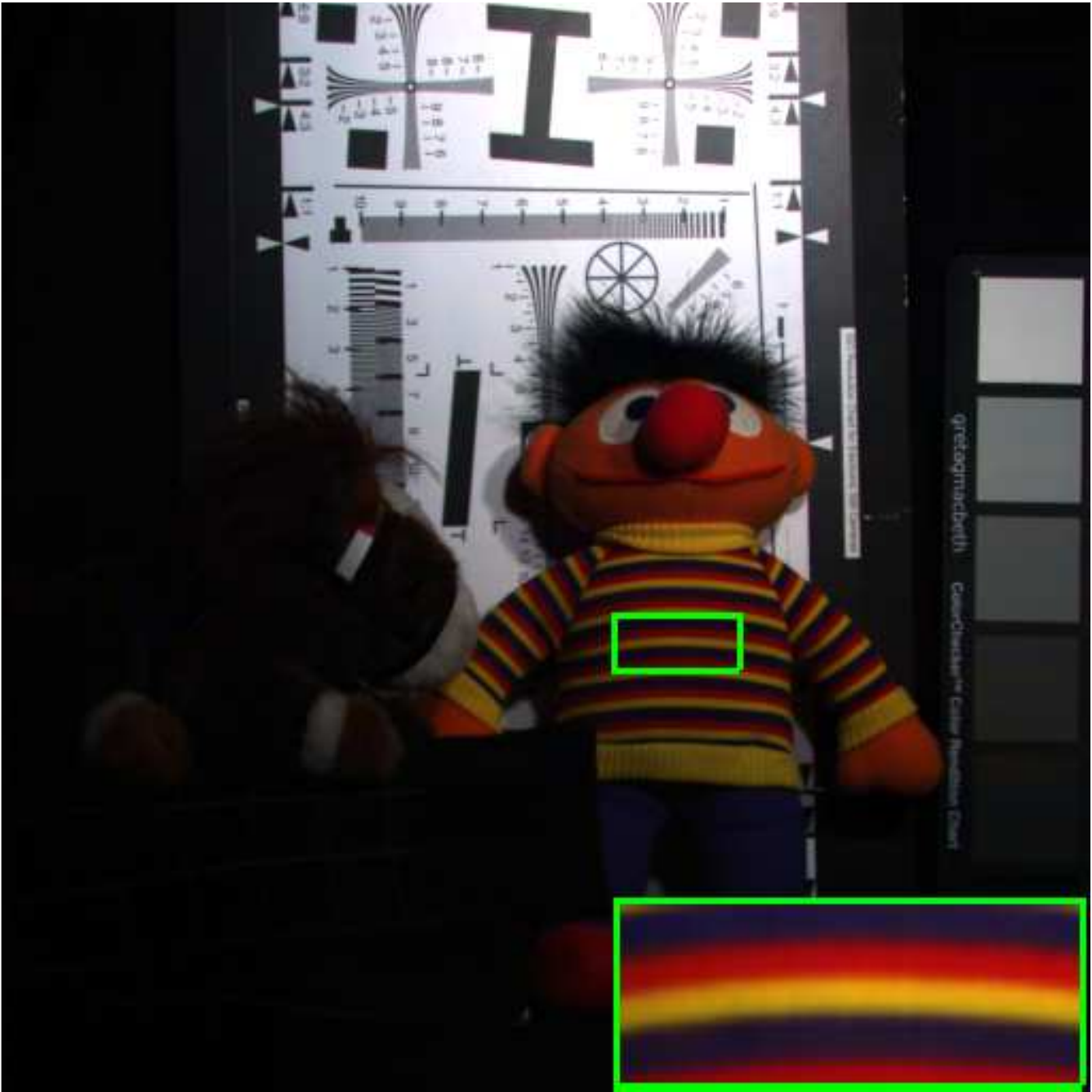} \\
    \includegraphics[width=\textwidth]{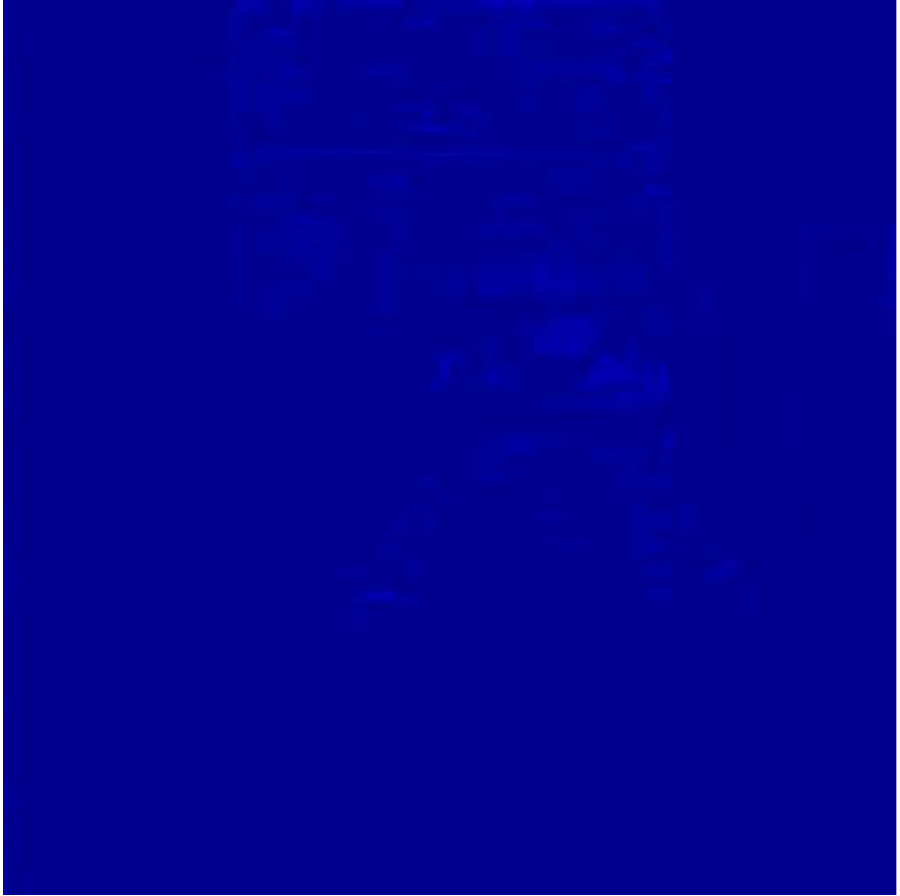} \\ \footnotesize(d) NCTRF
    \end{minipage}
    \begin{minipage}[t]{0.08\textwidth}\centering
     \includegraphics[width=\textwidth]{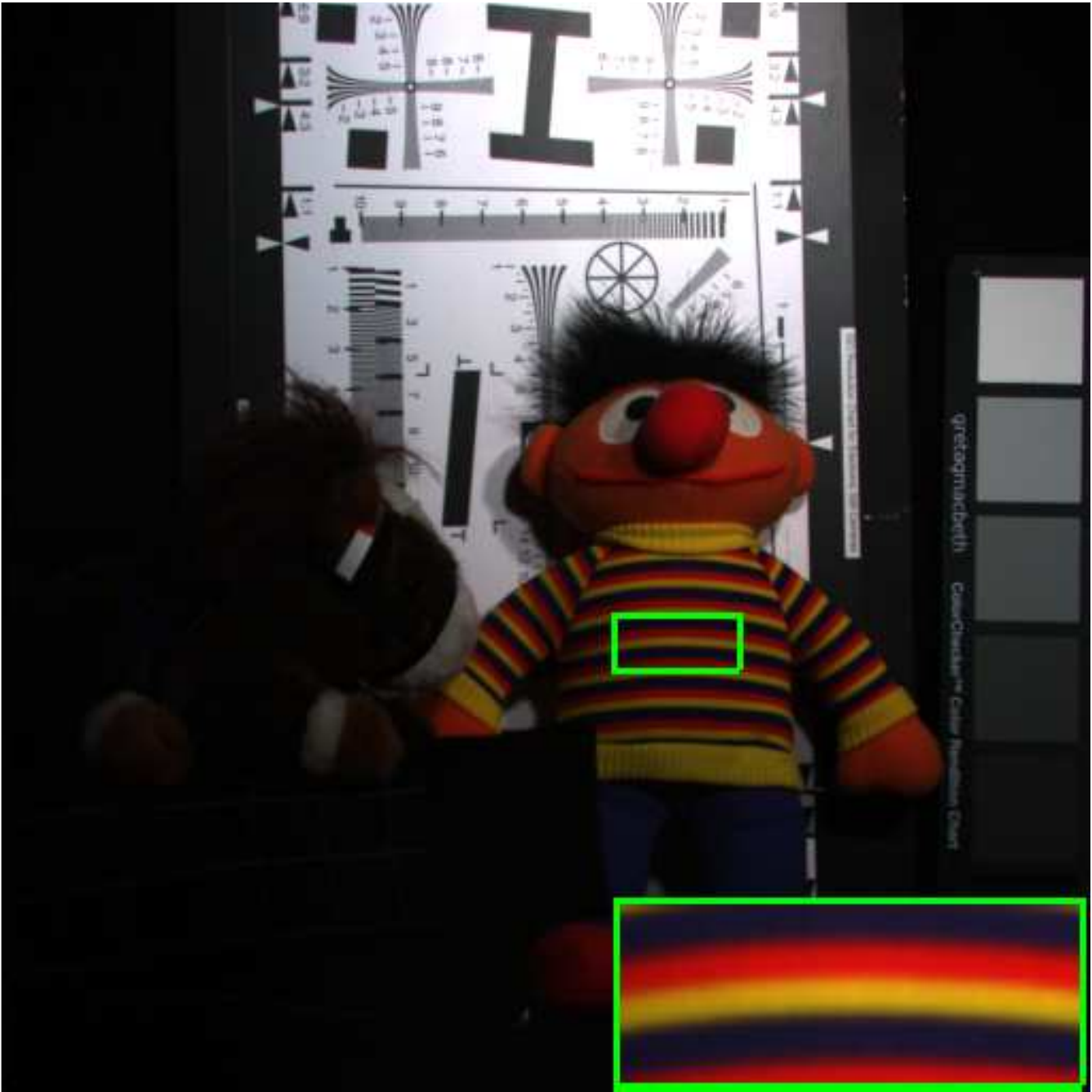} \\
    \includegraphics[width=\textwidth]{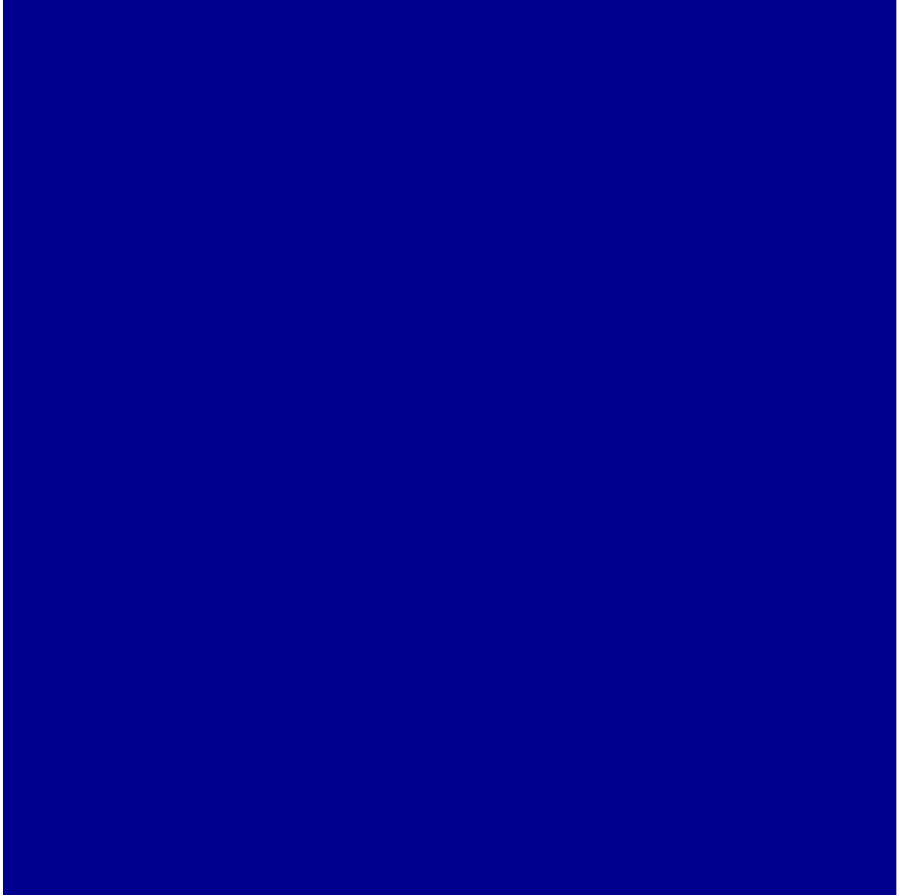} \\  \footnotesize (e) HR-HSI
    \end{minipage}
     \begin{minipage}{0.05\textwidth}\centering
  \includegraphics[width=0.95\textwidth,height=2.7\textwidth]{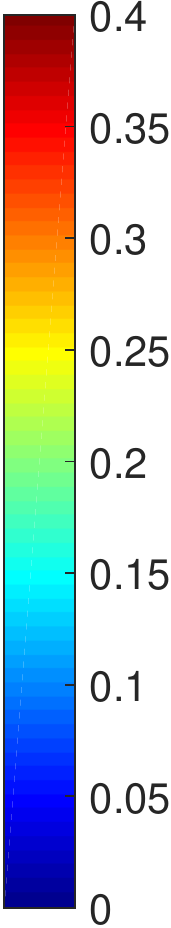}\centering \\
  \end{minipage} \\
\caption{HSR results of different methods with CAVE dataset at wavelength 460, 540 and 620 nm. The second and forth rows illustrate the difference images between HSR results and the ground-truth HR-HSI.}
\vspace{-7px}
\label{fig:CAVE_reconstruct}
\end{figure}

\noindent
\textbf{Comparison with deep learning.} Recently, deep learning related methods have also been introduced for the fusion of HSI and MSI~\cite{dian2018deep,nie2018deeply,qu2018unsupervised,xie2019multispectral,shi2018deep,Fu_2019_CVPR}. We take the unsupervised uSDN~\cite{qu2018unsupervised}
\footnote{\url{https://github.com/aicip/uSDN}}
and supervised MHF-net~\cite{xie2019multispectral}
\footnote{\url{https://github.com/XieQi2015/MHF-net}}
as state-of-the-art deep learning methods to compare with the proposed NCTRF method. As the same setting in~\cite{xie2019multispectral}, we adopt CAVE dataset
\footnote{\url{http://www1.cs.columbia.edu/CAVE/databases/}} for the experiments, with $20$ CAVE images for the training, and CAVE Toy and Balloons for the test. The spatial downsampling is the same as that in ~\cite{xie2019multispectral}, and the spectral downsampling matrix is the given spectral response matrix of Nikon D700~\cite{qu2018unsupervised}. The size of the simulated CAVE image is presented in Table~\ref{tab:datasize}. Table~\ref{tab:DL} presents the comparison between proposed method with two deep learning related methods. From the table, it can be observed that the proposed NCTRF achieves better quantitative evaluation results to uSDN. The proposed NCTRF can achieve higher PSNR values and lower SSIM values, compared to MHF-net. However, our proposed method needn't additional samples for training. Figure \ref{fig:CAVE_reconstruct} presents the visual results of different HSR methods, with related difference images between the HSR results and the ground truth HR-HSI CAVE images. From the difference images achieved by different methods, it can be observed that the proposed method can achieve better results compared to state-of-the-art deep learning methods.

\section{Conclusions}
In this paper, we propose a coupled tensor ring factorization (CTRF) model for the HSR. The proposed model inherits the advantage of coupled matrix and Tucker factorization, and is illustrated to better exploit the low-rank property of different HSI classes. Furthermore, we propose the NCTRF model by utilizing nuclear norm regularization of the third core tensor to exploit the global spectral low-rank property of the recovered HR-HSI. An efficient alternating iteration method has been proposed to optimize CTRF and NCTRF. The numerous experiments have demonstrated the advantage of the proposed methods compared to other tensor and deep learning methods. For future work, we plan to develop an automatic method to choose the TR \textit{rank}.
\section{Supplementary Material}
\subsubsection{Theorem~\ref{th:1}}
\begin{proof}
According to Proposition~\eqref{pr:2}, the circularly shifted tensor $\overleftarrow{\tensor{H}_{n-1}} \in \mathbb{R}^{I_{n}\times\cdots\times I_{N}\times I_1 \times{}\cdots \times I_{n-1}} $ can be expressed as
$$\overleftarrow{\tensor{H}_{n-1}}=\Phi(\{\tensor{G}^{(n)},\ldots,\tensor{G}^{(N)},\tensor{G}^{(1)},\ldots,\tensor{G}^{(n-1)}\}).$$
We merge the last $n-1$ core tensors using Proposition~\eqref{pr:1} and obtain the following

\begin{equation}
\tensor{H}_{<n>} = \tensor{G}^{(n)}_{(2)}\times (\mat{G}^{(n+1,\cdots,N,1,\cdots,n-1)}_{<2>})^{\top}.
\end{equation}

We record $\tensor{G}^{\bot} = (\mat{G}^{(n+1,\cdots,N,1,\cdots,n-1)}_{<2>})^{\top}$, according the property of matrix multiplication, the relation of the rank satisfies
\begin{equation}
\begin{split}
rank(\tensor{H}_{<n>})
& \leq
\min\{rank(\tensor{G}^{(n)}_{(2)}),
rank(\tensor{G}^{\bot})\}  \\
& \leq  rank(\tensor{G}^{(n)}_{(2)}).
\end{split}
\end{equation}
Since $\tensor{H}_{<n>}$ is the permutation of $\tensor{H}_{(n)}$ and $\tensor{G}^{(n)}_{<2>}$ is the permutation of $\tensor{G}^{(n)}_{(2)}$, we have the following:
\begin{equation}
\begin{split}
rank(\tensor{H}_{<n>}) = rank(\tensor{H}_{(n)})
&\leq rank(\tensor{G}^{(n)}_{<2>})= rank(\tensor{G}^{(n)}_{(2)}).
\end{split}
\end{equation}

\end{proof}

\ifCLASSOPTIONcaptionsoff
  \newpage
\fi
\bibliographystyle{IEEEtran}
\bibliography{egbib}

\end{document}